\documentclass[aps, showpacs,preprintnumbers, superscriptaddress, nofootinbibt, twocolumn]{revtex4-2}
\usepackage{amssymb}
\usepackage{afterpage}
\usepackage{graphicx}
\usepackage{makeidx}
\usepackage{amsfonts}
\usepackage{amsmath}
\usepackage[hyperindex,breaklinks]{hyperref}
\usepackage{eurosym}
\usepackage{hyperref}
\usepackage{color}
\usepackage{revsymb}
\usepackage[dvipsnames]{xcolor}
\usepackage{tensor}
\usepackage{xcolor}
\usepackage{eurosym}
\usepackage{amsmath}
\usepackage{bm}
\usepackage{indentfirst}
\usepackage{verbatim}
\usepackage{subcaption}
\usepackage{float}
\usepackage[utf8]{inputenc}

\hypersetup{
    colorlinks = true,
    linkcolor = blue,
    anchorcolor = blue,
    citecolor = blue,
    filecolor = blue,
    urlcolor = blue,
    breaklinks=true
   }

\def\be{\begin{equation}}
\def\ee{\end{equation}}
\def\bea{\begin{eqnarray}}
\def\eea{\end{eqnarray}}

\def\be{\begin{equation}}
\def\ee{\end{equation}}
\def\bea{\begin{eqnarray}}
\def\eea{\end{eqnarray}}

\begin{document}

\title{Warm inflation in Weyl geometric gravity}
\author{Runhua Huang}
\email[e-mail:]{huangrh27@mail2.sysu.edu.cn}
\affiliation{ School of Physics,
Sun Yat-Sen University, Guangzhou 510275, People’s Republic of China,}
\author{Tiberiu Harko}
\email{tiberiu.harko@aira.astro.ro}
\affiliation{Department of Physics, Babe\c s-Bolyai University, 1 Kog\u alniceanu Street, Cluj Napoca 400084, Romania,} 
\affiliation{Astronomical Observatory, 19 Cire\c silor Street, Cluj-Napoca 400487, Romania}
\author{Shi-Dong Liang}
\email{stslsd@mail.sysu.edu.cn}
\affiliation{ School of Physics,
Sun Yat-Sen University, Guangzhou 510275, People’s Republic of China,}
\author{Hong-Hao Zhang}
\email{zhh98@mail.sysu.edu.cn}
\affiliation{ School of Physics,
Sun Yat-Sen University, Guangzhou 510275, People’s Republic of China,}
\author{Lei Ming}
\email{minglei@scnu.edu.cn}
\affiliation {Key Laboratory of Atomic and Subatomic Structure and Quantum Control (Ministry of Education), Guangdong Basic Research Center of Excellence for Structure and Fundamental Interactions of Matter, School of Physics, South China Normal University, Guangzhou 510006, China} 
\affiliation {Guangdong Provincial Key Laboratory of Quantum Engineering and Quantum Materials, Guangdong-Hong Kong Joint Laboratory of Quantum Matter, South China Normal University, Guangzhou 510006, China}

\begin{abstract}
We investigate the warm inflationary scenario in the Weyl geometric gravity theory, in which the action is constructed by adding matter to the simplest conformally invariant gravitational action in Weyl geometry. The $\tilde{R}^2$ theory can be formulated equivalently as a linear theory supplemented by an additional scalar degree of freedom originating from higher-order curvature terms, with the equations of motion obtained via variational methods. We investigate the cosmological implications of the theory by considering the warm inflationary scenario of the early evolution of the Universe, in which radiation, the inflaton field, and the Weyl vector coexist. We consider the widely studied linear dissipation coefficient model along with a quartic potential, and investigate the influence of the Weyl vector term on the dynamics. We have performed numerical computations for different coupling models, and we have successfully developed a warm inflationary model in which the Universe transitions naturally from an inflationary epoch to a radiation-dominated era. The relevant cosmological observables have been calculated and compared with the latest observational constraints from the ACT data.
\end{abstract}

\date{\today}
\maketitle
\tableofcontents

{\ \hypersetup{linkcolor=blue} }
\section{Introduction}

Inflationary cosmology materialized first as a theory in \cite{1}, where the question of homogeneity and isotropy of space was answered by proposing a rapid expansion of the Universe in its early stages of existence. During this phase, the Universe underwent an exponential growth driven by a scalar field called inflaton, which is assumed to be homogenous and isotropic.
	
The theory of inflation has been supported by various observational evidences, such as the precise measurements of the cosmic microwave background radiation by satellite missions like the Cosmic Background Explorer (COBE) and the Wilkinson Microwave Anisotropy Probe (WMAP), as well as more recent observations by the Planck satellite \cite{COBE:1992syq, WMAP:2003ivt, Planck:2018jri}.
	
In the context of inflationary theories, the warm inflation scenario presents itself as an alternative to the standard inflation model, by integrating thermal effects considered to emerge from the interaction between the inflaton and other fields \cite{2}. In contrast to the standard inflationary theory, which assumes that the inflaton field evolves in vacuum, warm inflation takes into consideration the presence of thermal radiation, leading to dissipative processes \cite{3, Berera:1998px}. The inflation dynamics are therefore remodeled as the scalar field is coupled to a thermal bath of particles, allowing for energy transfer between the scalar field and its surrounding environment. This energy exchange leads to the dissipation of the inflaton's energy density, resulting in a slower, more gradual expansion of the Universe compared to the standard inflation. 

Since its proposal, the warm inflation model has been extensively studied, as reviewed in \cite{Berera:2023liv, Bastero-Gil:2009sdq, Kamali:2023lzq}. Warm inflation addresses and mitigates a number of long-standing challenges present in the cold inflation paradigm. The fine-tuning issue in cold inflation stems from the need for an exceptionally flat inflaton potential ($\epsilon_V, \eta_V\ll 1$) to produce long enough inflation, any small variation in the initial conditions would result in an insufficient number of e-folds being generated. In contrast, warm inflation intrinsically incorporates the dissipative ratio $Q$, which reshapes the effective slow-roll parameters ($\epsilon_H\approx\frac{\epsilon_V}{1+Q}$) and permits steeper potentials to support inflation \cite{Bastero-Gil:2016qru, Bastero-Gil:2019gao, Bartrum:2013fia, Berghaus:2019whh, Kamali:2023lzq}. Therefore, in the warm inflation scenario, inflation can still proceed even when the inflaton mass is of the order of the Hubble parameter ($m\sim H$) and the slow-roll parameter is of order unity ($\eta_V\sim\mathcal{O}(1)$). This provides a resolution to the so-called ``$\eta$ problem" without requiring extreme fine-tuning of the initial conditions. In cold inflation models, simple potentials (e.g., quadratic or quartic) often require inflaton field values exceeding the Planck scale ($\delta\phi>M_\text{pl}$), conflicting with effective field theory consistency and quantum gravity compatibility. While in warm inflation, in the strong dissipation regime ($Q\gg1$), field excursions are suppressed through dissipative friction, keeping them sub-Planckian \cite{Motaharfar:2018zyb, Berera:1999ws, Das:2020xmh}, and thus satisfying the so-called ``Swampland conjectures"\cite{Ooguri:2006in, Obied:2018sgi}.

 In addition, the presence of dissipation in warm inflation results in a reduced energy scale for the inflaton compared to cold inflation under the same potential, which in turn suppresses the tensor-to-scalar ratio. Consequently, a class of primordial potentials ruled out in cold inflation by CMB constraints becomes consistent with observations in the warm inflation framework \cite{Kamali:2023lzq, Benetti:2016jhf}.
	 
The ``graceful exit" problem was one of the most important original motivations for moving beyond cold inflationary models. In cold inflation, the Universe stays exponentially cold ($T_r\ll H$) throughout almost the entire inflationary phase. Inflation ends abruptly when the slow-roll conditions are violated ($\epsilon_H\approx 1$), and the inflaton oscillates around its potential minimum. Without interactions, the Universe would remain cold and empty, unable to smoothly transition into the hot, radiation‑dominated era required for the Hot Big Bang, and hence inconsistent with nucleosynthesis and established cosmological observations. Thus, a separate reheating stage is necessary, in which the inflaton decays into particles to produce radiation and reheat the Universe. 

One of the key features of warm inflation is that it can potentially lead to a graceful exit from inflation without the need for additional mechanisms. Warm inflation incorporates the interactions between the inflaton field and other particle degrees of freedom during the inflationary epoch, which are typically disregarded or considered negligible within the cold inflation paradigm. Inflaton field dissipates its energy to a thermal bath throughout inflation, via the dissipation term $\Gamma\dot{\psi}^2$ \cite{2,3}. Compared to cold inflation, the exit problem in warm inflation is more complex. The graceful exit of warm inflation hinges on three independent choices: the inflationary potential, the form of the dissipative coefficient, and whether the dynamics lie in the weak‑ or strong dissipation regime. The graceful exit problem under different forms of the dissipation coefficient $\Gamma$ is discussed in \cite{Das:2020lut, Saha:2025tln, Alhallak:2022szt}.

In warm inflation, where the temperature exceeds the Hubble rate ($T_r>H$), the CMB anisotropies are sourced by the classical thermal fluctuations of the radiation bath, as opposed to the quantum-mechanical origin in standard cold inflation, representing a fundamentally different mechanism for generating primordial perturbations. The first detailed numerical treatment of the evolution of scalar perturbations is provided in \cite{4}, by performing a full numerical integration of the complete set of cosmological scalar‑perturbation equations, beyond earlier analytic approximations. Subsequently, article \cite{Graham:2009bf} extended the computation of the scalar spectral index to the case of a temperature-dependent dissipation coefficient $\Gamma$. Different forms of the dissipation coefficient affect the scalar curvature $\mathcal{P}_R$ through a correction function $G(Q)$. Some recent study \cite{Montefalcone:2023pvh, Rodrigues:2025neh} present numerical code designed to solve for the perturbations’ equations in different warm inflation models. They also provide an analytic fit for $G(Q)$ for the cases of cubic, linear and inverse dissipation coefficient.

In recent studies \cite{Sharif:2015vda, Ghorui:2025vpu, Shiravand:2024ayw, Yeasmin:2022bqq, Yeasmin:2024yel}, warm inflation within the framework of modified gravity has begun to be explored. Warm inflation in the framework of modified Gauss-Bonnet gravity, specifically $f(G)$ gravity, is explored in \cite{Sharif:2015vda}, by using a self-interacting scalar field in the flat FLRW Universe. The field equations are construct under slow-roll approximation, and scalar and tensor power spectra are calculated. Warm inflation driven by $f(Q)$ dark energy is studied in \cite{Ghorui:2025vpu}. The paper computes the models corresponding to two different forms of $f(Q)$ functions and constrains the corresponding observables using data from Planck and BICEP/Keck. Article \cite{Shiravand:2024ayw} has investigated the warm inflationary scenario within the context of linear $f(Q, T)$ gravity, coupled with both the inflaton field and the radiation field. Using the slow‑roll approximation, the paper obtains both the scalar and the tensor power spectra, and consequently the expression for the scalar and tensor perturbations.

Recently, the Atacama Cosmology Telescope (ACT) released its latest measurements of the CMB data \cite{ACT:2025fju, ACT:2025tim, AtacamaCosmologyTelescope:2025vnj}. CMB lensing data from ACT and Planck are combined with baryon acoustic oscillation measurements from DESI DR1. The spectral index is found to be $n_s=0.9743\pm0.034$. This value is slightly higher than the Planck 2018 final result of $n_s = 0.9649\pm0.0044$ \cite{BICEP:2021xfz}, indicating stronger scale invariance in the scalar spectrum. Several recent papers have examined inflation models in light of the ACT data \cite{Okada:2025nyd, Lynker:2025wyc, Zharov:2025evb, Ellis:2025zrf, Okada:2025lpl, Haque:2025uga, Wang:2025cpp, Balkenhol:2025wms, Berera:2025vsu, Chakraborty:2025yms, Chakraborty:2025jof}, with particular focus on the higher scalar spectral index. A comprehensive update on primordial inflationary parameters using the most recent datasets is provided in \cite{Balkenhol:2025wms}. The work also provide a visualisation tools for custom $r-n_s$ plots against any potential. Article \cite{Wang:2025cpp} have discussed the inflation in framework of Weyl gravity. They propose a cold inflationary scenario from the Weyl scale-invariant gravity theory dominated by the higher-order curvatures. The inflationary effect originates from both the geometric contribution of higher‑order curvature terms and the dynamics of a physical inflaton field. Warm inflation models consistent with ACT data have been proposed in \cite{Berera:2025vsu, Chakraborty:2025yms, Chakraborty:2025jof}. The paper \cite{Berera:2025vsu} disfavors simple single-field cold inflation models from particle physics due to severe fine-tuning and eta-problem issues, and instead proposes warm inflation with natural dissipative couplings protected by mirror and $Z_4$ symmetries. The article demonstrates that the model naturally matches the observed nearly scale-invariant spectrum through strong dissipation regimes, by numerical computations. The paper \cite{Chakraborty:2025yms} investigates fibre inflationary models, which originally analyzed in cold inflation, by embedding them into the warm inflation framework. They demonstrate that various potentials remain viable in both strong and weak dissipative regimes, successfully matching the recent ACT observations. The paper \cite{Chakraborty:2025jof} examines $\alpha$-attractor models of inflation in the context of warm inflation, demonstrating that the attractor behavior persists in warm setups, allowing these models to remain compatible with recent CMB observations.

The aim of this work is to formulate a warm inflation model within the framework of the  Weyl geometric gravitational theory with action $R^2$ by introducing both an inflaton field $\psi$ and radiation $\rho_r$. By introducing an additional scalar field $\phi$, we reformulate the theory as a linear gravitational theory coupled to an extra scalar degree of freedom. This geometric scalar field, originating from higher‑order curvature terms, plays a dynamic role during the inflationary epoch. We demonstrate that within Weyl $\tilde{R}^2$ geometry, the coupling of the Weyl vector field is indispensable for the consistent inclusion of matter terms.

To investigate the behavior of the early universe, we adopt the flat Friedmann-Lemaitre-Robertson-Walker (FLRW) metric and derive the corresponding Friedmann equations. We derive the key equation of warm inflation—the energy balance equation—and examine warm inflation under different coupling models. For each model, we employ numerical methods to simulate the dynamical evolution of the Universe during the inflationary epoch and use the latest ACT data to constrain the model parameters. In this work, we employ numerical methods to show that warm inflation within the Weyl gravity framework can also provide a graceful exit mechanism while satisfying observational constraints.

The present paper is organized as follows. In Section \ref{2}, we provide a brief review of Weyl geometry, present our theoretical model, and derive the relevant field equations. In Section \ref{3}, we begin with a brief introduction to the standard warm inflation scenario, then establish our own warm inflation model in the flat FLRW Universe, and finally derive the corresponding Friedmann equations. In Section \ref{4}, we conduct numerical simulations for each warm inflation model, investigate the inflationary dynamics of the Universe under different model parameters, compare the model predictions with CMB observational data, and ultimately derive constraints on the model parameters. In Section \ref{5}, we discuss and summarize the results of this work and propose possible directions for future research. 

In this work, we use spacetime metrics with $\left(-,+,+,+\right)$ signature, and we use natural units for speed of light, reduced Planck constant and Boltzmann constant $c=\hbar=k_B=1$.
	
\section{Weyl geometry, action, and gravitational field equations}\label{2}

In the present Section we review the fundamentals of the Weyl geometric gravity. As a first step we briefly introduce the basic concepts of Weyl geometry, to be used in the sequel. Then we present the action and the field equations in the scalar-tensor representation of the conformally invariant Weyl action. The problem of the addition of the matter in the Lagrangian and the trace condition is also discussed. 

\subsection{Quick introduction to Weyl geometry}

The standard approach for obtaining the gravitational field equations, and to describe the gravitational interaction
	is based on the variational approach, and with the use of the Hilbert-Einstein Lagrangian, defined in a
	(pseudo-)Riemannian geometry, and given by
	\begin{equation}  \label{E}
		L_{H-E}=-\frac{1}{2}\left(M^2_p R +2\,M^2_p \Lambda\right)\sqrt{-g}
	\end{equation}
	where $M_p$ denotes the Planck mass, $R$ is the Ricci scalar constructed with the help of the metric tensor $g_{\mu \nu}$, and $\Lambda$ is the cosmological
	constant. 

In the following we assume that $L_{H-E}$ represents a low energy approximation of a spontaneously
	broken phase of a locally scale (conformally) invariant action \cite{Gh1,Gh2,Gh3,Gh4}. 
	The conformal symmetry is introduced by requiring the invariance of the natural laws and of the geometric and physical quantities 
	with respect to the group of transformations
	\begin{eqnarray}  \label{WS}
		\hat g_{\mu\nu}(x)&=&\Sigma^n(x) \,g_{\mu\nu}(x),\;\; \sqrt{\hat g(x)}%
		=\Sigma^{2 n}(x) \sqrt{g}(x),  \notag \\[5pt]
		 \hat \phi (x)&=& \Sigma^{-n/2}(x) \phi (x), \;\;
		\;\;\hat\psi (x)=\Sigma^{-3n/4}(x)\,\psi (x),
	\end{eqnarray}
	where $g=\vert\det g_{\mu\nu}\vert$, $\Sigma(x)>0$, and $\phi$ and $\psi$
	describe real bosonic or fermions fields. In the following, without
	any loss of generality, we take the Weyl charge as $n=1$.
	
	A natural way to enact the requirement of the conformal invariance is to apply
	the mathematical formalism of Weyl geometry. In its most general formulation  Weyl
	geometry can be interpreted as the classes of equivalence $(g_{\alpha \beta
	},\omega _{\sigma }$) of the metric $g_{\alpha \beta }$ and the Weyl vector gauge
	field ($\omega _{\sigma }$), which are related via the Weyl gauge conformal
	transformation rules (\ref{WS}). To the transformations (\ref{WS}) one must also add the
	transformation law of the Weyl vector $\omega _{\mu }$, given by
	\begin{equation}
		\hat{\omega}_{\mu }(x)=\omega _{\mu }(x)-\frac{1}{\alpha }\,\partial _{\mu }\ln
		\Sigma (x),
	\end{equation}%
	where by $\alpha $ we have denoted the Weyl gauge coupling parameter, a dimensionless constant. A central
	characteristic of the Weyl geometry is its non-metric character, since in this geometry the covariant
	derivative of the metric tensor does not vanish identically. An important geometric quantity in Weyl geometry is the non-metricity $Q_{\mu \alpha \beta}$, determined by
	the existence of a non-vanishing $\omega _{\mu }$, and defined according to
	\begin{equation}
		\tilde{\nabla}_{\mu }g_{\alpha \beta }=-\alpha \omega _{\mu }g_{\alpha \beta
		}\equiv Q_{\mu \alpha \beta}.  \label{nm}
	\end{equation}%
	From Eq.~(\ref{nm}) one finds the connection $\tilde{\Gamma}$ of the Weyl
	geometry as having the form
	\begin{eqnarray}
		\tilde{\Gamma}_{\mu \nu }^{\lambda } &=&\Gamma _{\mu \nu }^{\lambda }+\frac{1%
		}{2}\alpha \big[\delta _{\mu }^{\lambda }\,\,\omega _{\nu }+\delta _{\nu
		}^{\lambda }\,\,\omega _{\mu }-g_{\mu \nu }\,\omega ^{\lambda }\big]=\Gamma
		_{\mu \nu }^{\lambda }+\Xi _{\mu \nu }^{\lambda },  \notag  \label{tGamma}
		\\
		&&
	\end{eqnarray}%
	where $\Gamma _{\mu \nu }^{\lambda }$ is the Levi-Civita connection, given
	by its usual definition, $\Gamma _{\mu \nu }^{\alpha }(g)=(1/2)g^{\alpha
		\lambda }(\partial _{\mu }g_{\lambda \nu }+\partial _{\nu }g_{\lambda \mu
	}-\partial _{\lambda }g_{\mu \nu })$, and 
\be
\Xi _{\mu \nu }^{\lambda
	}=\frac{\alpha}{2}\Big[\delta _{\mu }^{\lambda }\,\,\omega _{\nu }+\delta _{\nu
	}^{\lambda }\,\,\omega _{\mu }-g_{\mu \nu }\,\omega ^{\lambda }\Big],
\ee
	respectively. Hence, in Weyl geometry, the covariant derivative of an arbitrary  vector $V_\mu$
	is obtained as
	\begin{eqnarray}
		\tilde{\nabla}_{\mu }V _{\nu } &=&\frac{\partial V _{\nu }}{%
			\partial x^{\nu }}-\tilde{\Gamma}_{\mu \nu }^{\lambda }V _{\lambda }=%
		\frac{\partial V _{\nu }}{\partial x^{\nu }}-\Gamma _{\mu \nu
		}^{\lambda }V _{\lambda }-\Xi _{\mu \nu }^{\lambda }V _{\lambda }
		\nonumber\\
		&=&\nabla _{\mu }V _{\nu }-\Xi _{\mu \nu }^{\lambda }V _{\lambda
		},
	\end{eqnarray}
	where $\nabla_\mu$ denotes the covariant derivative constructed with the help of the standard Levi-Civita connection.

	The contraction of the connection coefficients are obtained as $\tilde{\Gamma}%
	_{\mu \nu }^{\nu }=\tilde{\Gamma}_{\mu }$, and $\Gamma _{\mu \nu }^{\nu
	}=\Gamma _{\mu }$, respectively. The Weyl vector can thus be represented as
	\begin{equation}
		\omega _{\mu }=\frac{1}{2}\left( \tilde{\Gamma}_{\mu }-\Gamma _{\mu }\right).
	\end{equation}
	Therefore, $\omega _{\mu }$ can be interpreted as describing the deviation of the trace of the Weyl
	connection from the Levi-Civita connection. Since $%
	\omega _{\mu }$ is a component of the Weyl connection $\tilde{\Gamma}$, the trace expression shows that it has a
	purely geometric origin. The covariant divergence of a vector $V^\lambda$ in the
	Weyl geometry is obtained as
	\begin{equation}
		\tilde{\nabla}_{\lambda }V ^{\lambda }=\nabla _{\lambda }V
		^{\lambda }+2\alpha \omega _{\lambda }V ^{\lambda },
	\end{equation}

	The Weyl connection $\tilde{\Gamma}$ has the important property of being invariant with respect to the
	 set of transformations (\ref{WS}). In the particular case in which  $\omega _{\mu
	}$ decouples, $\omega _{\mu }=0$, or if it can be represented in a pure gauge form, as the gradient of a scalar function, then the particular geometry obtained in this way 
	is called the Weyl integrable geometry. Similarly to the general Weyl geometry, it is also a metric geometry.
	
	One can associate to the Weyl vector $\omega_\mu$ the field strength $F_{\mu\nu}$, defined as
	\begin{equation}  \label{W}
		\tilde{F}_{\mu\nu} = \tilde{\nabla}_{\mu} \omega_{\nu} - \tilde{\nabla}%
		_{\nu} \omega_{\mu} = \nabla _{\mu}\omega _\nu-\nabla _\nu \omega _\mu.
	\end{equation}
	
	The tensor curvatures $\tilde R^\lambda_{\mu\nu\sigma}$, $\tilde R_{\mu\nu}$ and the curvature scalar $\tilde{R}$ of Weyl geometry can be obtained with the help of the Weyl connection $\tilde\Gamma$. Their definitions
	are identical to the ones used in the Riemannian geometry, but with the Riemannian $\Gamma $
	replaced with the Weylian $\tilde\Gamma$. Hence, we obtain
	\begin{eqnarray}
		\tilde R^\lambda_{\mu\nu\sigma}&=& \partial_\nu
		\tilde\Gamma^\lambda_{\mu\sigma} -\partial_\sigma
		\tilde\Gamma^\lambda_{\mu\nu}
		+\tilde\Gamma^\lambda_{\nu\rho}\,\tilde\Gamma_{\mu\sigma}^\rho
		-\tilde\Gamma_{\sigma\rho}^\lambda\,\tilde\Gamma_{\mu\nu}^\rho,  \notag \\
		\tilde R_{\mu\nu}&=&\tilde R^\lambda_{\mu\lambda\sigma}, \tilde
		R=g^{\mu\sigma}\,\tilde R_{\mu\sigma}.
	\end{eqnarray}
	
	The full expressions of the Weyl curvature tensors and of the Weyl scalar are  
	\begin{eqnarray}  \label{tRmunu}
		\tilde R_{\mu\nu}&=&R_{\mu\nu} +\frac {1}{2}\alpha \left(\nabla_\mu \omega
		_\nu-3\,\nabla_\nu \omega _\mu - g_{\mu\nu}\,\nabla_\lambda \omega
		^\lambda\right)  \notag \\
		&&+\frac{1}{2} \alpha ^2 (\omega _\mu \omega _\nu -g_{\mu\nu}\,\omega
		_\lambda \omega ^\lambda),
	\end{eqnarray}
	\begin{equation}
		\hspace{-3.8cm}\tilde R_{\mu\nu}-\tilde R_{\nu\mu}=2 \alpha F_{\mu\nu},
	\end{equation}
	\begin{equation}  \label{tR}
		\hspace{-1.5cm}\tilde R= R-3 \,\alpha\,\nabla_\lambda\omega ^\lambda-\frac{3%
		}{2}\alpha ^2 \omega _\lambda \omega ^\lambda,
	\end{equation}
	where $R_{\mu \nu}$ and $R$ are the Ricci curvature tensor, and the Ricci
	scalar curvature defined in the Riemann geometry. We would like to point out that the right hand sides of the
	above equations are defined in the Riemannian geometry, with the covariant
	derivative given by the standard expression $\nabla_\mu V
	^\lambda=\partial_\mu V ^\lambda+\Gamma^\lambda_{\mu\rho}\,V ^\rho$.
	
	Under the conformal transformations (\ref{WS}), $\tilde R$ transforms covariantly, so that $\hat{\tilde R}=(1/\Sigma^n)\,\tilde R$.
	
\subsection{Action and field equations}

We will start our investigations with the following action of the Weyl geometric gravity theory \cite{Gh1,Gh2,Gh3,Gh4}, which also includes the  matter sector,
\begin{equation}\label{action1}
	S=\int\left(\frac{1}{4!\xi^2}\tilde{R}^2-\frac{1}{4}F^{\mu\nu}F_{\mu\nu}+\mathcal{L}_{\text{eff}}\right)\sqrt{-g}\mathrm{d}x^4.
\end{equation}
Action \eqref{action1} represents the simplest conformally invariant gravitational action, defined in the Weyl geometry framework, together with an effective matter term $\mathcal{L}_{\text{eff}}$. In an inflationary framework the effective matter Lagrangian $\mathcal{L}_{\text{eff}}$ contains the inflaton field and radiation, but we will further demonstrate that in order to maintain the conformal invariance of the action it must also incorporate the Weyl vector. 

In the Weyl geometric gravity theory, the action (\ref{action1}) can be linearized through the following procedure. We introduce first an auxiliary scalar field $\phi$, and make the substitution $\tilde{R}^2\longrightarrow2\phi^2\tilde{R}-\phi^4$. Then the action \eqref{action1} becomes
\begin{equation}\label{action2}
	S=\int\left(\frac{2}{4!\xi^2}\phi^2\tilde{R}-\frac{1}{4!\xi^2}\phi^4-\frac{1}{4}F^{\mu\nu}F_{\mu\nu}+\mathcal{L}_{\text{eff}}\right)\sqrt{-g}\mathrm{d}x^4.
\end{equation}
Variation of the action \eqref{action2} with respect to the scalar field $\phi$ gives 
\begin{equation}\label{scalar}
	\tilde{R}=\phi^2.
\end{equation}
Substituting Eq.~\eqref{scalar} back into the action~\eqref{action2} we reobtain \eqref{action1}. Hence, action \eqref{action2} is equivalent to the initial one. The higher-order curvature theory is thereby equivalent to a linear theory plus an extra scalar field.

By replacing $\tilde{R}$ in Eq.~\eqref{action2} with Eq.~\eqref{tR}, we obtain the action of the Weyl geometric theory defined in the Riemannian space, and given by
\begin{align}\label{action}
	S=\int\left(\frac{2}{4!\xi^2}\phi^2\left(R-3\alpha\nabla_\lambda\omega ^\lambda-\frac{3}{2}\alpha^2\omega_\lambda\omega^\lambda\right)\right.\\\nonumber
	\left.-\frac{1}{4!\xi^2}\phi^4-\frac{1}{4}F^{\mu\nu}F_{\mu\nu}+\mathcal{L}_{\text{eff}}\right)\sqrt{-g}\mathrm{d}x^4.
\end{align}
The variation of the action \eqref{action} with respect to the metric tensor gives the field equation
\begin{align}\label{field}
	&\phi^2\left( R_{\mu\nu}-\frac{1}{2}Rg_{\mu \nu }\right)
	+\left(g_{\mu\nu}\square-\nabla_\mu\nabla_\nu\right)\phi^2\nonumber\\
	&+\frac{3}{2}\alpha^2\phi^2\left(\frac{1}{2}
	\omega _\rho\omega ^\rho g_{\mu \nu }-\omega _{\mu }\omega _{\nu }\right)+\frac{1}{4}\phi^4g_{\mu \nu }\nonumber\\
	&+\frac{3}{2}\alpha\left(\omega _{\mu }\nabla _{\nu }+\omega _{\nu }\nabla _{\mu }-g_{\mu \nu
	}\omega^\rho\nabla_\rho\right)\phi^2\nonumber \nonumber\\
	&-6\xi^2\left(F_\mu^\rho F_{\nu\rho}-\frac{1}{4}g_{\mu \nu }F_{\alpha
		\beta }F^{\alpha \beta }\right)-6\xi^2T^{(\text{eff})}_{\mu\nu}=0,
\end{align}
where the energy-momentum tensor $T^{(\text{eff})}_{\mu\nu}$ of the effective matter Lagrangian $\mathcal{L}_{\text{eff}}$ is defined as 
\begin{equation}\label{Tmn}
	T^{(\text{eff})}_{\mu\nu}\equiv-\frac{2}{\sqrt{-g}}\frac{\delta\left(\sqrt{-g}\mathcal{L}_{\text{eff}}\right)}{\delta g^{\mu\nu}}=g_{\mu\nu}\mathcal{L}_\text{eff}-2\frac{\partial\mathcal{L}_\text{eff}}{\partial g^{\mu\nu}}.
\end{equation}

The variation of the action \eqref{action} with respect to the Weyl vector $\omega^\mu$ yields the equation of motion of $\omega^\mu$, given by
\begin{equation}\label{vector}
	-4\xi^2\nabla ^{\nu }F_{\mu\nu}+\alpha\nabla_\mu\phi^2-\alpha^2\phi^2\omega_\mu+4\xi^2\frac{\partial\mathcal{L}_\text{eff}}{\partial\omega^\mu}=0.
\end{equation}

From the trace of Eq.~\eqref{field} we obtain the equation
\begin{equation}\label{trace}
	-R\phi^2+3\square\phi^2+\frac{3}{2}\alpha^2\phi^2\omega_\rho\omega^\rho+\phi^4-3\alpha\omega^\rho\nabla_\rho\phi^2-6\xi^2T^{(\text{eff})}=0,
\end{equation}
where $T^{(\text{eff})}\equiv g^{\mu\nu}T^{(\text{eff})}_{\mu\nu}$. The scalar Eq.~\eqref{scalar} can be explicitly written as
\begin{equation}\label{scalar1}
	R-3\alpha\nabla_\mu\omega^\mu-\frac{3}{2}\alpha^2\omega_\mu\omega^\mu=\phi^2.
\end{equation}

Combining Eqs. \eqref{trace} and \eqref{scalar1}, it follows that
\begin{equation}\label{scalar2}
	\square\phi^2-\alpha\nabla_\mu\left(\phi^2\omega^\mu\right)-2\xi^2T^{(\text{eff})}=0.
\end{equation}
Applying the operator $\nabla^\mu$ to both sides of Eq.~\eqref{vector}, we obtain
\begin{equation}\label{vector2}
	-4\xi^2\nabla^\mu\nabla ^{\nu }F_{\mu\nu}+\alpha\square\phi^2-\alpha^2\nabla^{\mu }\left(\phi^2\omega_\mu\right)+4\xi^2\nabla^\mu\frac{\partial\mathcal{L}_\text{eff}}{\partial\omega^\mu}=0.
\end{equation}
Due to the antisymmetric nature of $F_{\mu\nu}$, and of the Bianchi identity, the term $\nabla^\mu\nabla ^{\nu }F_{\mu\nu}$ identically vanishes. The comparison of Eq.~\eqref{vector2} with Eq.~ \eqref{scalar2} yields the following relation
\begin{equation}\label{consis}
	T^{(\text{eff})}+\frac{2}{\alpha}\nabla^\mu\frac{\partial\mathcal{L}_\text{eff}}{\partial\omega^\mu}=0.
\end{equation}

In the absence of matter ($\mathcal{L}_{\text{eff}}=0$), the model reduces to the simplest Weyl $\tilde{R}^2$ geometric gravity theory. Then Eq.~\eqref{consis} is trivially satisfied, demonstrating the consistency of the model. When matter with a non-zero trace of the energy-momentum tensor is added to the model, $\mathcal{L}_{\text{eff}}$ must also include the Weyl vector to satisfy Eq.~\eqref{consis}. Thus, $\mathcal{L}_{\text{eff}}$ can be assumed to consist of the matter Lagrangian $\mathcal{L}_m$, and the Lagrangian $\mathcal{L}_\omega$ associated with the Weyl vector. Thus, the effective Lagrangian is given by 
\begin{equation}\label{effLa}
	\mathcal{L}_{\text{eff}}=\mathcal{L}_m+\mathcal{L}_\omega.
\end{equation}
Generally, $\mathcal{L}_\omega$ can depend on the inflaton field $\psi$, thereby introducing a non-minimal coupling between the inflaton and the Weyl vector field. 

\section{Warm inflation in Weyl geometric gravity}\label{3}

In the present Section we first briefly review the standard warm inflationary scenario as developed in the framework of general relativity. Then the generalized Friedmann equations for the cosmological evolution in the framework of Weyl geometric gravity are derived. Finally, we formulate the warm inflationary scenario in the $\tilde{R}^2$ theory in the presence of the conformally coupled matter.

\subsection{Standard warm inflationary scenario}

The requirement from General Relativity for realising an inflationary phase is only the dominance of vacuum energy density. So it does not rule out the possibility that a substantial amount of radiation energy may remain during inflation. Thus, the most general picture of inflation accommodates a radiation energy density component. In warm inflationary scenario, it postulates that the inflaton remains continuously coupled to the radiation field during the inflationary epoch, thereby sustaining a thermal bath at a finite temperature. This mechanism eliminates the need for a distinct reheating phase and allows a natural transition to a radiation dominated Universe after inflation. In this section, we briefly review the dynamics of warm inflation within the framework of General Relativity.

In warm inflationary scenario, there are two kinds of matter components: the radiation and the inflaton field. In the flat FLRW Universe,  the Friedmann equations are given by
\begin{equation}\label{fr1}
	3H^2=8\pi G\left(\rho_r+\rho_\psi\right),
\end{equation}
\begin{equation}\label{fr2}
	2\dot{H}+3H^2=-8\pi G\left(p_r+p_\psi\right),
\end{equation}
where $G$ is Newton’s gravitational constant, $H$ is the Hubble function, and a dot over a variable denotes its time derivative. $\rho_r, \rho_\psi$ are energy densities of radiation and inflaton, respectively, and $p_r, p_\psi$ are pressures of radiation and inflaton. The density and pressure of the scalar field are
\begin{equation}\label{exp1}
	\rho_\psi=\frac{1}{2}\dot{\psi}^2+V(\psi),
\end{equation}
\begin{equation}\label{exp2}
	p_\psi=\frac{1}{2}\dot{\psi}^2-V(\psi),
\end{equation}
where $V(\psi)$ is scalar potential. When vacuum energy dominates over kinetic energy and radiation ($V(\psi)\gg\frac{1}{2}\dot{\psi}^2,\rho_r$), the Hubble function vary slowly, i.e., $\dot{H}\approx0, H\approx\text{const}$. The Universe undergoes exponential expansion. The evolution of the inflaton field is governed by the equation \cite{Bastero-Gil:2009sdq, Berera:2008ar, Berera:2023liv},
\begin{equation}\label{scal}
	\ddot{\psi}+\left(3H+\Gamma\right)\dot{\psi}+V'(\psi)=0,
\end{equation}
where $V'(\psi)\equiv\frac{\mathrm{d}V}{\mathrm{d}\psi}$. Eqs. \eqref{fr1}, \eqref{fr2} and \eqref{scal} set the background dynamics of warm inflation. Considering Eqs. \eqref{exp1} and \eqref{exp2}, Eq. \eqref{scal} is equivalent to
\begin{equation}\label{scaevol}
	\dot{\rho_\psi}+3H\left(\rho_\psi+p_\psi\right)=-\Gamma\dot{\psi}^2.
\end{equation}
The term on the right-hand side of Eq. \eqref{scaevol} indicates the dissipation of inflaton's energy. According to energy conservation, this dissipated energy must be transferred to radiation, thereby generating radiation during the inflationary epoch. Eqs. \eqref{fr1}, \eqref{fr2} and \eqref{scaevol} yield the energy balance equation for radiation, given by
\begin{equation}\label{rad}
	\dot{\rho_r}=-4H\rho_r+\Gamma\dot{\psi}^2.
\end{equation}
This equation is central to the study of warm inflation. The first term on the right-hand side constitutes an energy sink, thereby depleting radiation, whereas the second term serves as a source that supplies energy. The first term on the right-hand side decays away any initial radiation, while the second term maintains the radiation at a nonzero value. Thus at large times, compared to the Hubble time, the radiation in the Universe becomes independent of initial conditions and depends only on the rate at which the source is producing radiation.

In standard analyses of warm inflation, the radiation bath is assumed to be in near-thermal equilibrium. This allows for the definition of a temperature $T_r$, which is related to the energy density of radiation. One can write \cite{Ghosh:2025vza}
\begin{equation}\label{temp}
	\rho_r=C_*T_r^4,
\end{equation}
where $C_*$ is the Stefan–Boltzmann constant and is defined by $C_*\equiv\frac{\pi^2}{30}g_*$. $g_*$ is the number of degrees of freedom of the radiation field. For $Tr\gtrsim300\text{GeV}$, all the species of the Standard Model are in equilibrium, $g_*\simeq106.75$ \cite{Riotto:2018pcx, Husdal:2016haj}.

$\Gamma$ in Eq. \eqref{scal} is the dissipative coefficient, which characterizes the interaction between the inflaton field and all other fields. $\Gamma$ is treated as a phenomenological function. It takes a generic form of \cite{Kumar:2024hju, Das:2020lut}
\begin{equation}\label{key}
	\Gamma=\gamma T_r^a\psi^bM^{1-a-b},
\end{equation}
where $\gamma$ is a dimensionless coefficient and $M$ is an appropriate mass scale. The specific forms of the dissipative coefficient $\Gamma$ in warm inflation models are chosen based on microphysical derivations from quantum field theory interactions. The dissipation arises from the inflaton $\psi$ coupling to other fields (bosonic or fermionic), which form or interact with a thermal radiation bath. The exact temperature $T_r$ and field $\psi$ dependence emerges from finite-temperature QFT calculations, considering the exact interaction structure. Warm inflation with different temperature-dependent dissipation coefficients are widely studied in \cite{Moss:2006gt, Ito:2025lcg, Berghaus:2025dqi, Bastero-Gil:2012akf, Santos:2024pix, Sahu:2025osa, Zhang:2009ge, Bastero-Gil:2010dgy, ORamos:2025uqs, Bastero-Gil:2006ahd}. Among these, the cubic ($\Gamma\propto\frac{T_r^3}{\psi^2}$), linear ($\Gamma\propto T_r$), and inverse ($\Gamma\propto\frac{1}{T_r}$) forms of the dissipative coefficient have received the most extensive study. The cubic form typically arises in the high-temperature regime of supersymmetric or two-stage decay models \cite{Moss:2006gt, Ito:2025lcg, Bastero-Gil:2010dgy}, where the inflaton dissipates into heavy fields that decay into light radiation without large thermal corrections to the potential.

The cubic dissipation coefficient is a classical choice for realizing strong‑dissipation warm inflation. The linear form commonly arises from fermionic couplings or axion-like interactions with gauge fields \cite{Berghaus:2025dqi, ORamos:2025uqs, Ito:2025lcg, Cheng:2024uvn}, enabling viable strong or weak dissipative regimes while protecting the potential from large thermal corrections. Linear dissipation is increasingly favored in recent literature for its stability, SM compatibility, and observational fit compared to cubic coefficients. The inverse form typically arises in the low-temperature regime of supersymmetric \cite{Bastero-Gil:2012akf}, where thermal corrections to particle masses lead to suppressed dissipation. Inverse dissipation is less favored in recent observational comparisons compared to linear or cubic. Furthermore, numerous studies have explored composite or more sophisticated forms of the dissipative coefficient \cite{Sahu:2025osa, Santos:2024pix, Zhang:2013yr}. The detailed procedure for deriving the form of the dissipative coefficient from field theory is discussed in depth in \cite{Kamali:2023lzq, Berera:2023liv}.

In warm inflation, the presence of dissipation affects the background dynamics of the inflaton field as well as its perturbations. When $T_r>H$, thermal fluctuations can dominate quantum density fluctuations. The dimensionless primordial scalar curvature power spectrum with the thermal effects is obtained as \cite{Bastero-Gil:2016qru, Berera:2018tfc}
\begin{equation}\label{PR}
	\mathcal{P}_R = \left( \frac{H_*^2}{2\pi \dot{\psi}_*} \right)^2\mathcal{F},
\end{equation}
where $\mathcal{F}$ is the modified factor, and its specific form is given by
\begin{equation}\label{PRF}
	\mathcal{F}=\left(
	1 + 2n_{*} + \frac{T_{r*}}{H_*} \frac{2\sqrt{3}\pi Q_*}{\sqrt{3 + 4\pi Q_*}}
	\right)
	G(Q_*),
\end{equation}
where $Q\equiv\frac{\Gamma}{3H}$, and subscript $*$ correspond to the quantity evaluated at horizon crossing. At full thermal equilibrium between the inflaton and radiation, $n_*=n_\text{BE}=\frac{1}{e^{H_*/T_{r*}}-1}$, where $n_\text{BE}$ is the Bose-Einstein distribution. If the inflaton field does not’t thermalize with the radiation bath, it is customary to take $n_*=0$. For some intermediate cases between these two extremes (i.e., where $n_*$ lies between $0$ and $n_\text{BE}$), see \cite{Bastero-Gil:2017yzb}. The $G(Q_*)$ factor in Eq. \eqref{PRF} appears in the power spectrum as a signature of the coupling of the inflaton field with the radiation bath, and this growth factor can only be evaluated by numerically solving the full set of perturbation equations of warm inflation. The functional form of $G(Q_*)$ primarily depends on the form of the dissipative coefficient and has only weak dependence on the form of the potential.

Due to the weakness of gravitational interaction, the tensors modes do not’t get affected by the presence of the radiation bath
as there are no direct couplings between them, and thus, the tensor power spectrum remains the same as in cold inflation, given by \cite{Kumar:2024hju, Kitabayashi:2025jzh}
\begin{equation}\label{PT}
	\mathcal{P}_T=\frac{2H^2}{\pi^2M^2_p}
\end{equation}
where $M_p\equiv\frac{1}{\sqrt{8\pi G}}$ is the reduced Planck mass. The tensor-to-scalar ratio $r$ and scalar index $n_s$ are defined as
\begin{equation}\label{ratio}
	r=\frac{\mathcal{P}_T}{\mathcal{P}_R},
\end{equation}
\begin{equation}\label{ns}
	n_s-1=\frac{\mathrm{d}(\mathrm{ln}\mathcal{P}_R)}{\mathrm{d}(\ln k)}.
\end{equation}

\subsection{The generalized Friedmann equations}
We investigate warm inflationary models in the flat FLRW Universe, with the metric given by
\begin{equation}\label{FLRW}
	\mathrm{d}s^2=-\mathrm{d}t^2+a^2\left(t\right)\left(\mathrm{d}x^2+\mathrm{d}y^2+\mathrm{d}z^2\right),
\end{equation}
where $a(t)$ represents the scale factor. Due to the homogeneity and isotropy of the Universe, the Weyl vector field $\omega_\mu$ can possess only a time component, which can be expressed as:
\begin{equation}\label{Weylvector}
	\omega_\mu=\left(\omega(t),0,0,0\right).
\end{equation}
Thus, the field strength $F_{\mu\nu}$ of the Weyl vector $\omega_\mu$ vanish in the FLRW Universe. In this study,  without loss of generality, we adopt $\alpha=1$. Moreover, we model the radiation as an ideal fluid, characterized by its energy density $\rho_r$ and pressure $p_r$, with the equation of state $\rho_r=3p_r$. The total energy density and pressure are given by
\begin{equation}\label{re}
	\rho_\text{eff}=\rho_r+\rho_\psi+\rho_\omega,
\end{equation}
\begin{equation}\label{pe}
	p_\text{eff}=p_r+p_\psi+p_\omega.
\end{equation}
$\rho_r$ and $\rho_\psi$ are energy densities of radiation and inflaton, respectively. $\rho_\omega$ is contributed  by $\mathcal{L}_\omega$. $\rho_\omega$ and $p_\omega$ are related to $\mathcal{L}_\omega$ through Eq. \eqref{Tmn}, given by
\begin{equation}\label{rw}
	\rho_\omega=-\mathcal{L}_\omega-2\omega^2\frac{\partial\mathcal{L}_\omega}{\partial\omega^2},
\end{equation}
\begin{equation}\label{pw}
	p_\omega=\mathcal{L}_\omega.
\end{equation}
where $\frac{\partial\mathcal{L}_\omega}{\partial\omega^2}\equiv\frac{\partial\mathcal{L}_\omega}{\partial\left(\omega_\mu\omega^\mu\right)}$.

Considering the $(00)$ and $(ii)$ components of Eq. \eqref{field} yields the first and second generalized Friedmann equations, respectively, given by
\begin{equation}\label{fir}
	3H^2=-6H\frac{\dot{\phi}}{\phi}+\frac{1}{4}\phi^2+\frac{3}{4}\omega^2-3\omega\frac{\dot{\phi}}{\phi}+\frac{6\xi^2}{\phi^2}\rho_\text{eff},
\end{equation}
\begin{align}\label{sec}
	2\dot{H}+3H^2=
	&-2\frac{\ddot{\phi}}{\phi}-2\frac{\dot{\phi}^2}{\phi^2}-4H\frac{\dot{\phi}}{\phi}+\frac{1}{4}\phi^2\nonumber\\
	&-\frac{3}{4}\omega^2+3\omega\frac{\dot{\phi}}{\phi}-\frac{6\xi^2}{\phi^2}p_\text{eff}.
\end{align}
From Eqs. \eqref{fir} and \eqref{sec}, we obtain the following relation for the derivative of the Hubble function
\begin{equation}\label{Hdot}	\dot{H}=-\frac{\ddot{\phi}}{\phi}-\frac{\dot{\phi}^2}{\phi^2}+H\frac{\dot{\phi}}{\phi}-\frac{3}{4}\omega^2+3\omega\frac{\dot{\phi}}{\phi}-\frac{3\xi^2}{\phi^2}\left(\rho_\text{eff}+p_\text{eff}\right).
\end{equation}

The vector field equation Eq. \eqref{vector} becomes 
\begin{equation}\label{vec}
	\omega=2\frac{\dot{\phi}}{\phi}+\frac{8\xi^2\omega}{\phi^2}\frac{\partial\mathcal{L}_\omega}{\partial\omega^2}.
\end{equation}
The scalar equation Eq. \eqref{scalar1} becomes
\begin{align}\label{sca}
6\left(\dot{H}+2H^2\right)+3\left(\dot{\omega}+3H\omega\right)+\frac{3}{2}\omega^2-\phi^2=0.
\end{align}
From Eqs. \eqref{fir} and \eqref{Hdot}, we obtain the expression for deceleration parameter $q$, given by
\begin{widetext}
\begin{align}\label{key}	q\equiv-1-\frac{\dot{H}}{H^2}=\frac{12\xi^2\left(\rho_\text{eff}+3p_\text{eff}\right)-\phi^4+6\phi^2\omega^2+12\dot{\phi}^2+12\phi\left(\left(H-2\omega\right)\dot{\phi}+\ddot{\phi}\right)}{24\xi^2\rho_\text{eff}+\phi\left(\phi^3+3\phi\omega^2-12\left(2H+\omega\right)\dot{\phi}\right)}.
\end{align}
\end{widetext}

\subsection{Warm inflation in Weyl $\tilde{R}^2$ gravity}

The energy balance equation is derived by combining Eqs. \eqref{fir}, \eqref{sec}, \eqref{vec} and \eqref{sca}, in a similar way to the standard model. The derivation is complex, but straightforward. A step-by-step calculation is detailed in appendix \ref{A}. The energy balance equation is given by
\begin{equation}\label{energy}
	\dot{\rho}_m+3H\left(\rho_m+p_m\right)=\dot{\psi}\frac{\partial\mathcal{L}_\omega}{\partial\psi},
\end{equation}
where we have denoted $\rho_m\equiv\rho_r+\rho_\psi, p_m\equiv p_r+p_\psi$. Analogy to standard model, Eq. \eqref{energy} can split into
\begin{equation}\label{radiation}
	\dot{\rho_r}+4H\rho_r=\Gamma\dot{\psi}^2,
\end{equation}
\begin{equation}\label{inflaton}
	\ddot{\psi}+\left(3H+\Gamma\right)\dot{\psi}+V'(\psi)=\frac{\partial\mathcal{L}_\omega}{\partial\psi}.
\end{equation}
Similar to standard warm inflation, the $\Gamma$ term serves as the crucial link between radiation and the inflaton. It acts as a source term for radiation production while simultaneously dissipating the inflaton's energy. Furthermore, in our model, the $\mathcal{L}_\omega$ term introduces a novel coupling between the inflaton and the Weyl vector field. Depending on its sign, $\mathcal{L}_\omega$ can either source or dissipate the energy of the inflaton field.  In the following, we will investigate the impact of the $\mathcal{L}_\omega$ term on the behavior of the inflationary system.

In this study, we consider the dissipation parameter and the inflaton potential of the following forms:
\begin{equation}\label{key}
	\Gamma=\gamma T_r,
\end{equation}
\begin{equation}\label{key}
	V(\psi)=\kappa\psi^4,
\end{equation}
where $\gamma$ and $\kappa$ are dimensionless coefficients. In this work, we assume that the inflaton field is not in thermal equilibrium with radiation, and therefore $n_*=0$ in Eq. \eqref{PRF}. Different choices of $n_*$ can influence the specific form of $G(Q)$, and this issue is investigated in detail in \cite{Rodrigues:2025neh}. For a linearly $T_r$-dependent dissipation coefficient, function $G(Q)$ can be well approximated by \cite{Benetti:2016jhf, AlHallak:2022haa, Kamali:2023lzq}
\begin{equation}\label{key}
	G(Q)\simeq1+0.335Q^{1.364}+0.0185Q^{2.315}.
\end{equation}
This formula holds for $Q\lesssim50$. General and more accurate results can be found in \cite{Rodrigues:2025neh}.

In warm inflation, the tensor-to-scalar ratio $r$ and scalar index $n_s$ will be modified. In this work, we assume that the extra scalar degree of freedom $\phi$ and the Weyl vector $\omega$ have very small pertubations compared with those of the inflaton field $\psi$. Consequently, these additional degrees of freedom affect only the background dynamics and do not source the curvature perturbation at leading order. Therefore, we consider Eq. \eqref{PR} from the classical model to remain valid. According to Eqs. \eqref{PR}, \eqref{PT}, \eqref{ratio} and \eqref{ns}, $r$ and $n_s$ are given by \cite{Kitabayashi:2025jzh, AlHallak:2022haa, Cheng:2024uvn}
\begin{equation}\label{key}
	r=\frac{16\epsilon_H}{1+Q}\mathcal{F}^{-1},
\end{equation}
\begin{equation}\label{key}
	n_s-1=\frac{1}{H\mathcal{P}_R}\frac{\mathrm{d}\mathcal{P}_R}{\mathrm{d}t},
\end{equation}
where $\epsilon_H\equiv-\frac{\dot{H}}{H^2}$ is the slow-roll parameter. Inflation ends when $\epsilon_H$ reach 1. The Tensor-to-scalar ratio is suppressed by factor $\mathcal{F}$ and dissipation coefficient $1+Q$. In the limit $T_r\rightarrow0$ and $Q\rightarrow0$, the relation in cold inflation $r=16\epsilon_H$ is recovered.

The latest result on CMB data was released from Atacama Cosmology Telescope (ACT) \cite{ACT:2025fju, ACT:2025tim}. These results, combined with those of the DESI, produce a new constrain on the scalar index, given by
\begin{equation}\label{nscons}
	n_s=0.9743\pm0.0034.
\end{equation}
This result differs from the Planck result by at least $2\sigma$. In this work, we proposed the model consistent with the latest constraints on the scalar spectral index, while also satisfying the Planck bounds on the tensor-to-scalar ratio, given by \cite{BICEP:2021xfz}
\begin{equation}\label{ratiocons}
	r<0.036.
\end{equation}

The number of e-folds $N$ before the end of inflation is defined as
\begin{equation}\label{key}
	N\equiv\int^{t_\text{end}}_tH\mathrm{d}t,
\end{equation}
where $t_\text{end}$ represents the moment inflation ends. \textbf{In this work, we normalize power spectrum $\mathcal{P}_R=2.1\times10^{-9}$, at the pivot point $N_*=60$.}

\section{Specific warm inflationary models }\label{4}

In the present Section, we will investigate specific models of the warm inflationary scenario in Weyl geometric gravity. To simplify the mathematical formalism for numerical computation, we introduce a set of dimensionless variables, defined as
\begin{eqnarray}\label{dimens}
	&h\equiv\frac{H}{H_0},\quad\tau\equiv H_0t,\quad f\equiv\frac{\phi}{H_0},\quad\Psi\equiv\frac{\psi}{H_0},\\\nonumber
	&w\equiv\frac{\omega}{H_0},\quad r_\text{rad}\equiv\frac{\rho_r}{H_0^4},\quad\theta_r\equiv\frac{T_r}{H_0},\nonumber
\end{eqnarray}
where $H_0$ is the initial value of the Hubble function.

\subsection{The minimal coupling model}

In this section, we investigate the minimal coupling model, i.e., $\frac{\partial\mathcal{L}_\omega}{\partial\psi}=0$. $\mathcal{L}_\omega$ takes the form
\begin{equation}\label{key}
	\mathcal{L}_\omega=-\eta H_0^{4-2n}\left(-\omega_\mu\omega^\mu\right)^n.
\end{equation}
In the minimal coupling model, Eq. \eqref{energy} becomes
\begin{equation}\label{energy1}
	\dot{\rho}_m+3H\left(\rho_m+p_m\right)=0.
\end{equation}
Despite the presence of both the Weyl vector $\omega_\mu$ and the scalar field $\phi$ originating from geometric curvature, the energy balance equation Eq. \eqref{energy1} retains the same form as in standard warm inflation. The total energy of the inflaton and radiation is conserved, coupling to the background spacetime exclusively through the Hubble parameter $H$. Nevertheless, $\omega_\mu$ and $\phi$ can still indirectly influence the inflationary process through their interaction with the background spacetime. Below, we demonstrate that even in the minimal coupling model, the $\mathcal{L}_\omega$ term still have a substantial impact on the inflationary dynamics.

According to the definition Eqs. \eqref{dimens}, the dimensionless equations of the system are obtained as
\begin{align}\label{1sec}
	2\dot{h}+3h^2=
	&-2\frac{\ddot{f}}{f}-4h\frac{\dot{f}}{f}+\frac{1}{4}f^2-2\frac{\dot{f}^2}{f^2}-\frac{3}{4}w^2+3w\frac{\dot{f}}{f}\nonumber\\
	&-\frac{6\xi^2}{f^2}\left(\frac{1}{3}r_\text{rad}+\frac{1}{2}\dot{\Psi}^2-\kappa \Psi^4-\eta w^{2n}\right),
\end{align}
\begin{equation}\label{1vecd}
	\dot{w}=2\frac{\ddot{f}}{f}-2\frac{\dot{f}^2}{f^2}+8\xi^2\eta n\left(2n-1\right)\frac{w^{2n-2}\dot{w}}{f^2}-16\xi^2n\eta\frac{\dot{f}w^{2n-1}}{f^3},
\end{equation}
\begin{align}\label{1sca}
	6\left(\dot{h}+2h^2\right)+3\left(\dot{w}+3hw\right)+\frac{3}{2}w^2-f^2=0,
\end{align}
\begin{equation}\label{1radiation}
	\dot{r}_\text{rad}+4hr_\text{rad}=\gamma\theta_r\dot{\Psi}^2,
\end{equation}
\begin{equation}\label{1inflaton}
	\ddot{\Psi}+\left(3h+\gamma\theta_r\right)\dot{\Psi}+4\kappa \Psi^3=0.
\end{equation}
Eqs. \eqref{1sec}, \eqref{1sca}, \eqref{1radiation}, \eqref{1inflaton} come from Eqs. \eqref{sec}, \eqref{sca}, \eqref{radiation} and \eqref{inflaton}, respectively. Eq. \eqref{1vecd} comes from the derivative of Eq. \eqref{vec}. Eqs. \eqref{fir} and \eqref{vec} give 
\begin{equation}\label{1fir}
	3h^2=-6h\frac{\dot{f}}{f}+\frac{1}{4}f^2-\frac{3}{4}w^2+\frac{6\xi^2}{f^2}\left(r_\text{rad}+\frac{1}{2}\dot{\Psi}^2+\kappa \Psi^4-\eta w^{2n}\right),
\end{equation}
\begin{equation}\label{1vec}
	w=2\frac{\dot{f}}{f}+8\xi^2n\eta\frac{w^{2n-1}}{f^2}.
\end{equation}
Eqs. \eqref{1sec}, \eqref{1vecd}, \eqref{1sca}, \eqref{1radiation} and \eqref{1inflaton} are solved numerically, while Eqs. \eqref{1fir} and \eqref{1vec} constraining the initial conditions.

\subsubsection{Numerical results}

We chose for the initial conditions the following values: $h(0)=1, r_\text{rad}(0)=1\times10^{-3}, f(0)=0.1, w(0)=0.1, \Psi(0)=180, \Psi'(0)=0$. The value of $f'(0)$ and $\xi$ are constrained by Eqs. \eqref{1fir} and \eqref{1vec}, for any given values of $n, \kappa, \gamma$ and $\eta$. 

\begin{figure}[t]
	\includegraphics[width=8.5cm,height=8.5cm]{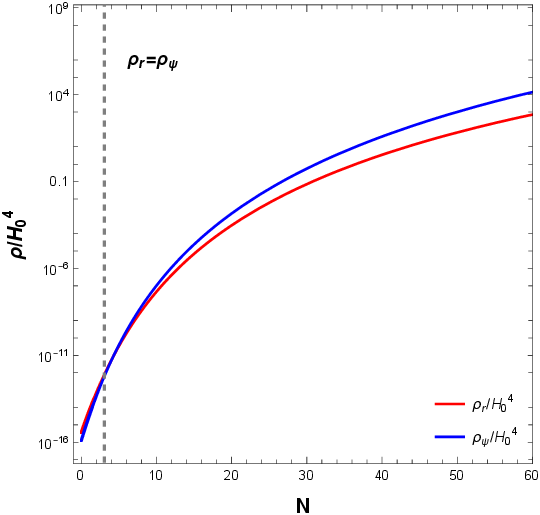}
	\caption{Evolution of the dimensionless radiation energy densities $\rho_r/H_0^4$ (red) and the dimensionless inflaton energy densities $\rho_\psi/H_0^4$ (blue) with number of e-folds $N$ in the model $\mathcal{L}_\omega=-\eta H_0^{4-2n}\left(-\omega_\mu\omega^\mu\right)^n$. $N$ evolves from a large value up to $N=0$ where inflation ends. Model parameters are: $\eta=0.5, \gamma=72, \kappa=0.08502, n=1.5$.}
	\label{1energy}
\end{figure}

As introduced earlier in Eq. \eqref{re}, the effective energy density $\rho_\text{eff}$ is composed of the radiation energy density $\rho_r$\, the inflaton energy density $\rho_\psi$, and the energy density $\rho_\omega$. In this model, for positive values of $\eta$, we obtain a negative $\rho_\omega$. Fig. \ref{1energy} shows the evolution of the dimensionless radiation energy densities $\rho_r/H_0^4$ (red) and the dimensionless inflaton energy densities $\rho_\psi/H_0^4$ with number of e-folds $N$. The black dashed line marks the moment radiation overtakes inflaton ($\rho_r=\rho_\psi$). During inflation, the energy density of the inflaton dominates the Universe, driving the inflationary expansion. Radiation and the inflaton field coexist and interact during inflation, which is the classic signature of warm inflation that distinguishes it from cold one. As the Universe expands, the energy density of the inflaton decreases more rapidly than that of radiation, and is overtaken by the latter at $N\approx4.2$. After radiation overtakes the inflaton, inflation continues to proceed for a certain duration until $N=0$.

\begin{figure}[t]
	\includegraphics[width=8.5cm,height=8.5cm]{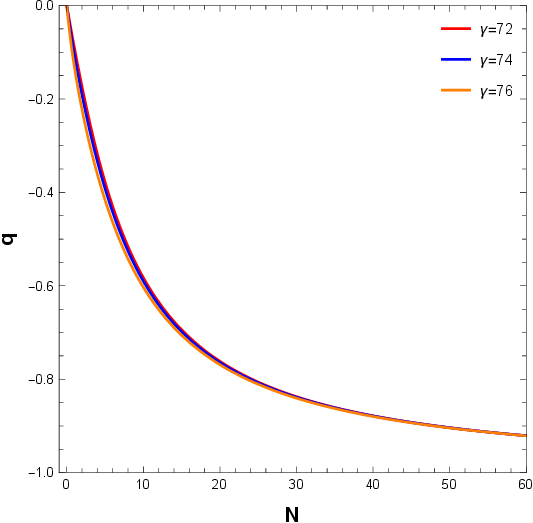}
	\caption{Variations as functions of number of e-folds $N$ of the deceleration parameter $q$ in the model $\mathcal{L}_\omega=-\eta H_0^{4-2n}\left(-\omega_\mu\omega^\mu\right)^n$ for different $\gamma$. $N$ evolves from a large value up to $N=0$ where inflation ends. The values of $\gamma$ and $\kappa$ are: $\gamma=72, \kappa=0.08502$ (red); $\gamma=74, \kappa=0.07966$ (blue); $\gamma=76, \kappa=0.07400$ (orange). Other parameters are: $\eta=0.5, n=1.5$.}
	\label{1q}
\end{figure}

The variations of deceleration parameter $q$ with respect to $N$ are represented in Fig. \ref{1q}. The deceleration parameter increases monotonically from negative values and eventually crosses zero. For different values of $\gamma$, the deceleration parameters exhibit nearly identical initial evolution, with deviations emerging only after a certain period. Across different parameter values, the models consistently depict the scenario of warm inflation, describing the Universe's transition from an inflationary epoch to a radiation-dominated era. 

\begin{figure*}[t]
	\begin{center}
		\begin{subfigure}{0.495\textwidth}
			\includegraphics[width=\textwidth, height=0.7\textwidth]{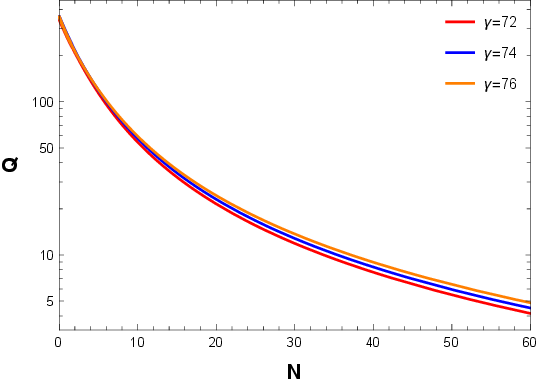}
		\end{subfigure}
		\hfill
		\begin{subfigure}{0.495\textwidth}
			\includegraphics[width=\textwidth, height=0.7\textwidth]{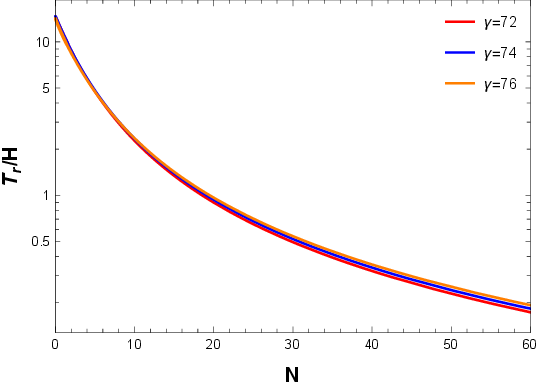}
		\end{subfigure}
	\end{center}
	\caption{Variations as functions of number of e-folds $N$ of dissipation ratio $Q$ (left panel) and thermal ratio $T_r/H$ (right panel) in the model $\mathcal{L}_\omega=-\eta H_0^{4-2n}\left(-\omega_\mu\omega^\mu\right)^n$ for different $\gamma$. $N$ evolves from a large value up to $N=0$ where inflation ends. The values of $\gamma$ and $\kappa$ are: $\gamma=72, \kappa=0.08502$ (red); $\gamma=74, \kappa=0.07966$ (blue); $\gamma=76, \kappa=0.07400$ (orange). Other parameters are: $\eta=0.5, n=1.5$.}
	\label{1QTH}
\end{figure*}

Fig. \ref{1QTH} shows the evolutions of dissipation ratio $Q$ and thermal ratio $T_r/H$ as a function of $N$ for different values of $\gamma$. The value of Q increases with $\gamma$. As $N$ decreases, $Q$ increases, and throughout this process $Q>1$, which is a characteristic of strong-dissipation warm inflation. Since $Q\propto T_r/H$ in the linear dissipation model, the two quantities exhibit similar evolution. The thermal ratio $T_r/H$ exceeds 1 only after $N\simeq20$. Therefore, at horizon crossing \textbf{($N_*=60$)}, thermal fluctuations do not significantly dominate over quantum ones.

\begin{figure*}[t]
	\begin{center}
	\begin{subfigure}{0.495\textwidth}
		\includegraphics[width=\textwidth, height=0.7\textwidth]{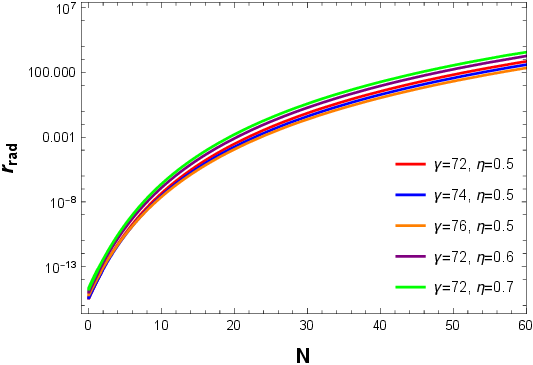}
	\end{subfigure}
	\hfill
	\begin{subfigure}{0.495\textwidth}
		\includegraphics[width=\textwidth, height=0.7\textwidth]{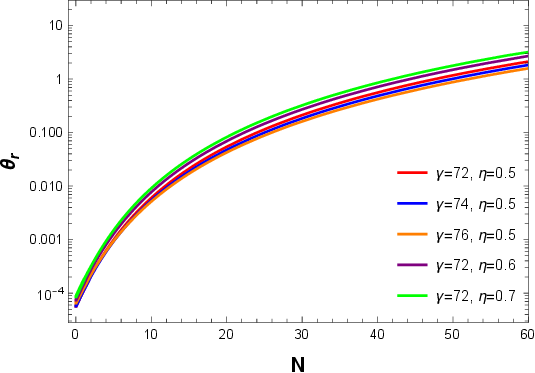}
	\end{subfigure}
\end{center}
	\caption{Variations as functions of number of e-folds $N$ of the dimensionless radiation density $r_\text{rad}$ (left panel) and of the dimensionless temperature $\theta_r$ (right panel) in the model $\mathcal{L}_\omega=-\eta H_0^{4-2n}\left(-\omega_\mu\omega^\mu\right)^n$ for different $\gamma, \eta$. $N$ evolves from a large value up to $N=0$ where inflation ends. The values of $\gamma, \eta$ are: $\gamma=72, \eta=0.5$ (red); $\gamma=74, \eta=0.5$ (blue); $\gamma=76, \eta=0.5$ (orange); $\gamma=72, \eta=0.6$ (purple); $\gamma=72, \eta=0.7$ (green).}
	\label{1rT}
\end{figure*}

The variations of the dimensionless radiation energy density $r_\text{rad}$ and the dimensionless temperature $\theta_r$ with respect to number of e-folds are represented in Fig. \ref{1rT}. Radiation is diluted during the violent inflationary expansion, but remains non-zero due to its interactions with the inflaton field. Both $\gamma$ and $\eta$ affect the radiation density during inflation,  with smaller $\gamma$ and larger $\eta$ corresponding to higher radiation density. Due to Eq. \eqref{temp}, the dimensionless temperature $\theta_r$ exhibits similar behavior to the radiation density. The temperature at the end of inflationary regime, i.e., the reheating temperature, $T_\text{reh}$, is one of the most important ingredients in the thermal history of the Universe. The exact value of the reheating temperature has not yet been known, and it lies in the extensive range $4\text{MeV}\lesssim T_\text{reh}\lesssim 10^{15}\text{GeV}$ \cite{Gim:2016uvv}. According to the definition of the dimensionless temperature $\theta_r$ in Eqs. \eqref{dimens}, the initial value of Hubble function $H_0$ and the reheating temperature $T_\text{reh}$ are related via

\begin{equation}\label{reheating}
	H_0=\frac{T_\text{reh}}{\theta_\text{end}}.
\end{equation}
As clearly showed in Fig. \ref{1rT}, the temperature at the end of inflation $\theta_\text{end}$ depend on $\gamma$ and $\eta$. The value of $\eta$ significantly affect $\theta_\text{end}$. The temperatures at the end of inflation $\theta_\text{end}$, and the corresponding values of $\gamma, \eta$ are given in Table~\ref{tab1temp}.

\begin{table}[htbp]
	\begin{center}
		\begin{tabular*}{0.5\columnwidth}{@{\extracolsep{\fill}}c c c@{\extracolsep{\fill}}}
			\hline
			\textbf{$\gamma$} & \textbf{$\eta$} & \textbf{$\theta_\text{end} (10^{-5})$} \\
			\hline
			$72$ & $0.5$ & $5.65$\\
			$74$ & $0.5$ & $5.69$\\
			$76$ & $0.5$ & $6.50$\\
			$72$ & $0.6$ & $7.36$\\
			$72$ & $0.7$ & $8.48$\\
			\hline
		\end{tabular*}
	\end{center}
	\caption{The temperature at the end of inflation $\theta_\text{end}$ for different values of $\gamma$ and $\eta$, in the model $\mathcal{L}_\omega=-\eta H_0^{4-2n}\left(-\omega_\mu\omega^\mu\right)^n$.}
	\label{tab1temp}
\end{table}

\begin{figure*}[t]
	\begin{center}
	\begin{subfigure}{0.495\textwidth}
		\includegraphics[width=\textwidth, height=0.7\textwidth]{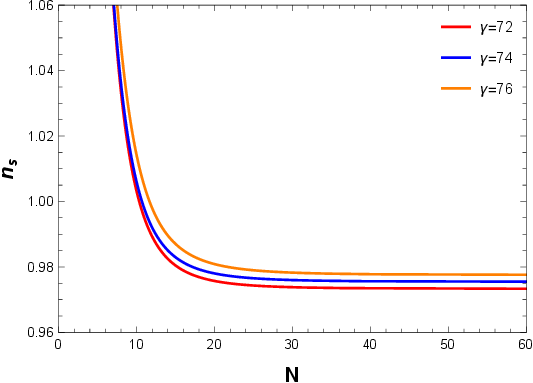}
	\end{subfigure}
	\hfill
	\begin{subfigure}{0.495\textwidth}
		\includegraphics[width=\textwidth, height=0.7\textwidth]{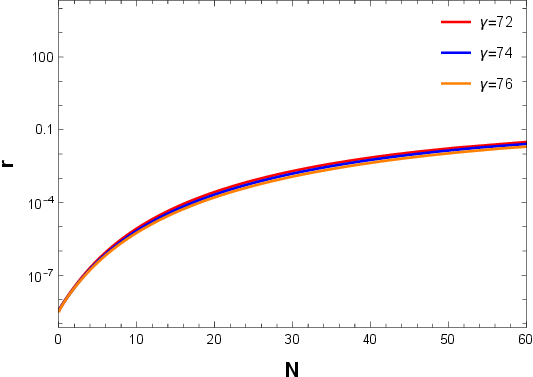}
	\end{subfigure}
\end{center}
	\caption{The values of scalar spectral index $n_s$ (left panel) and tensor-to-scalar ratio $r$ (right panel) with respect to number of e-folds $N$ for different parameters in the model $\mathcal{L}_\omega=-\eta H_0^{4-2n}\left(-\omega_\mu\omega^\mu\right)^n$. $N$ evolves from a large value up to $N=0$ where inflation ends.The values of $\gamma$ and $\kappa$ are: $\gamma=72, \kappa=0.08502$ (red); $\gamma=74, \kappa=0.07966$ (blue); $\gamma=76, \kappa=0.07400$ (orange). Other parameters are: $\eta=0.5, n=1.5$.}
	\label{1nsrN2}
\end{figure*}

Fig.~\ref{1nsrN2} shows the evolution curves of $r$ and $n_s$ with $N$ for different values of $\gamma$ at a fixed $\eta$. In the minimal coupling model, the values of $r$ and $n_s$ are significantly influenced by $\gamma$, while remaining largely insensitive to $\eta$. Larger values of $\gamma$ lead to a higher $n_s$ and a lower $r$. $n_s$ increases as $N$ decreases, while $r$ decreases with decreasing $N$.

\begin{figure}[t]
	\includegraphics[width=8.5cm, height=8.5cm]{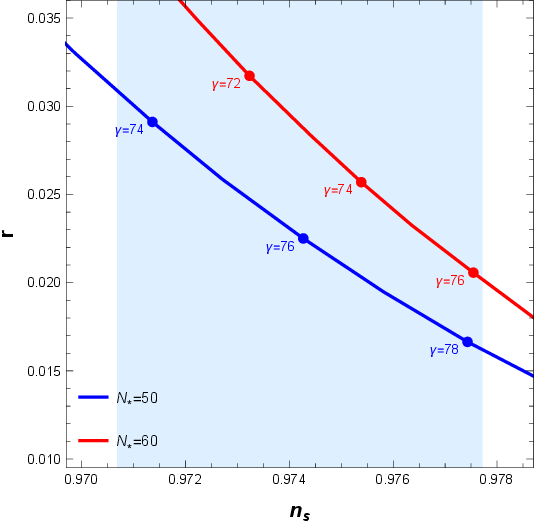}
	\caption{The tensor-to-scalar ratio $r$ against the scalar spectral index $n_s$ for different values of $N_*$: $N_*=50$ (blue), $N_*=60$ (red), in the model $\mathcal{L}_\omega=-\eta H_0^{4-2n}\left(-\omega_\mu\omega^\mu\right)^n$. On each curve, $\gamma$ increases from left to right. The blue region denotes the parameter space consistent with the constraints from Eqs. \eqref{nscons} and \eqref{ratiocons}. $N$ correspond to CMB pivot scale and the values of $\kappa$ are chosen to enforce $\mathcal{P}_R\simeq2.1\times10^{-9}$. Other parameters are: $\eta=0.5, n=1.5$.}
	\label{1rations}
\end{figure}

Fig.~\ref{1rations} plots the tensor-to-scalar ratio $r$ against the scalar spectral index $n_s$ for different values of $N_*$. The blue region denotes the parameter space consistent with the constraints from Eqs. \eqref{nscons} and \eqref{ratiocons}. On each curve, $\gamma$ increases from left to right. The values of $\kappa$ are chosen to enforce $\mathcal{P}_R\simeq2.1\times10^{-9}$ at $N_*=50$ (blue) and $N_*=60$ (red), respectively. $r$ decreases as $n_s$ increases. A suitable interval of parameter $\gamma$ can be identified for the chosen pivot point, within which the observables are consistent with the observational constraints. As $N_*$ increases, the curve shifts toward the upper‑right region. Beyond a certain maximum value of $N_*$, the curve no longer falls within the observationally allowed region. Table \ref{tab1N} presents the suitable ranges of $\gamma$ corresponding to different values of $N_*$. As $N_*$ increases, the allowed range of $\gamma$ shifts toward lower values.

\begin{table}[htbp]
	\begin{center}
	\begin{tabular*}{0.5\columnwidth}{@{\extracolsep{\fill}}c c@{\extracolsep{\fill}}}
		\hline
		\textbf{$N_*$} & \textbf{$\gamma$}  \\
		\hline
		$50$ & $73.56 - 78.17$ \\
		$60$ & $70.67 - 76.16$ \\
		\hline
	\end{tabular*}
\end{center}
	\caption{The suitable ranges of $\gamma$ corresponding to different values of $N_*$, in the model $\mathcal{L}_\omega=-\eta H_0^{4-2n}\left(-\omega_\mu\omega^\mu\right)^n$. Other parameters are: $\eta=0.5, n=1.5$.}
	\label{tab1N}
\end{table}

\subsection{The non-minimal coupling model}

In this Section, we consider a Lagrangian with the Weyl vector coupled with the inflaton, i.e., $\frac{\partial\mathcal{L}_\omega}{\partial\psi}\neq0$. $\mathcal{L}_\omega$ takes the form
\begin{equation}\label{key}
	\mathcal{L}_\omega=-\eta H_0^{4-2n-m}\psi^m\left(-\omega_\mu\omega^\mu\right)^n.
\end{equation}
The inflaton equation Eq. \eqref{inflaton} becomes
\begin{equation}\label{inflatonequation}
	\ddot{\psi}+\left(3H+\Gamma\right)\dot{\psi}+V'(\psi)=-\eta m H_0^{2-m}\psi^{m-1}\left(-\omega_\mu\omega^\mu\right)^n.
\end{equation}
Compared to the minimal coupling model, the $\mathcal{L}_\omega$ term in this scenario directly affects the dynamical evolution of the inflaton field. $\mathcal{L}_\omega$ represent an interaction of the inflaton field $\psi$ with the Weyl vector $\omega$. Eq. \eqref{inflatonequation} can be written as
\begin{equation}\label{}
	\dot{\rho}_\psi+3H(\rho_\psi+p_\psi)=-\Gamma\dot{\psi}^2-\eta m \dot{\psi}H_0^{2-m}\psi^{m-1}\left(-\omega_\mu\omega^\mu\right)^n.
\end{equation}
Given that $\eta>0$ and $\dot{\psi}<0$, the $\mathcal{L}_\omega$ term dissipates energy from inflaton when $m$ is negative, and sources energy into it when $m$ is positive. Both scenarios have significant effects on both the inflationary dynamics and the values of observables. In the following, we will study the specific implications of two different forms of coupling.

The dimensionless equations of the system are
\begin{align}\label{2sec}
	2\dot{h}+3h^2=
	&-2\frac{\ddot{f}}{f}-4h\frac{\dot{f}}{f}+\frac{1}{4}f^2-2\frac{\dot{f}^2}{f^2}-\frac{3}{4}w^2+3w\frac{\dot{f}}{f}\nonumber\\
	&-\frac{6\xi^2}{f^2}\left(\frac{1}{3}r_\text{rad}+\frac{1}{2}\dot{\Psi}^2-\kappa \Psi^4-\eta\Psi^m w^{2n}\right),
\end{align}
\begin{align}\label{2vecd}
	\dot{w}=
	&2\frac{\ddot{f}}{f}-2\frac{\dot{f}^2}{f^2}+8\xi^2\eta n\left(2n-1\right)\frac{w^{2n-2}\dot{w}}{f^2}\Psi^m\nonumber\\
	&-16\xi^2n\eta \frac{\dot{f}w^{2n-1}}{f^3}\Psi^m+8\xi^2n\eta m\frac{w^{2n-1}}{f^2}\Psi^{m-1}\dot{\Psi},
\end{align}
\begin{align}\label{2sca}
	6\left(\dot{h}+2h^2\right)+3\left(\dot{w}+3hw\right)+\frac{3}{2}w^2-f^2=0,
\end{align}
\begin{equation}\label{2radiation}
	\dot{r}_\text{rad}+4hr_\text{rad}=\gamma\theta_r\dot{\Psi}^2,
\end{equation}
\begin{equation}\label{2inflaton}
	\ddot{\Psi}+\left(3h+\gamma\theta_r\right)\dot{\Psi}+4\kappa \Psi^3=-\eta m\Psi^{m-1}w^{2n}.
\end{equation}
Eqs. \eqref{2sec}, \eqref{2sca}, \eqref{2radiation}, \eqref{2inflaton} come from Eqs. \eqref{sec}, \eqref{sca}, \eqref{radiation} and \eqref{inflaton}, respectively. Eq. \eqref{2vecd} comes from the derivative of Eq. \eqref{vec}. And the constraints on the initial conditions are
\begin{align}\label{2fir}
	3h^2=
	&-6h\frac{\dot{f}}{f}+\frac{1}{4}f^2-\frac{3}{4}w^{2n}\nonumber\\
	&+\frac{6\xi^2}{f^2}\left(r_\text{rad}+\frac{1}{2}\dot{\Psi}^2+\kappa \Psi^4-\eta\Psi^mw^{2n}\right),
\end{align}
\begin{equation}\label{2vec}
	w=2\frac{\dot{f}}{f}+8\xi^2n\eta\frac{w^{2n-1}}{f^2}\Psi^m,
\end{equation}
which come from Eqs. \eqref{fir} and \eqref{vec}. Eqs. \eqref{2sec}, \eqref{2vecd}, \eqref{2sca}, \eqref{2radiation} and \eqref{2inflaton} are solved numerically, while Eqs. \eqref{2fir} and \eqref{2vec} constraining the initial conditions.

\subsubsection{Numerical results}

We chose for the initial conditions the following values: $h(0)=1, r_\text{rad}(0)=1\times10^{-3}, f(0)=0.1, w(0)=0.1, \Psi(0)=180, \Psi'(0)=0$. The value of $f'(0)$ and $\xi$ are constrained by Eqs. \eqref{2fir} and \eqref{2vec}, for any given values of $\kappa, \gamma, \eta, n$ and $m$.

\begin{figure}[t]
	\includegraphics[width=8.cm, height=6.5cm]{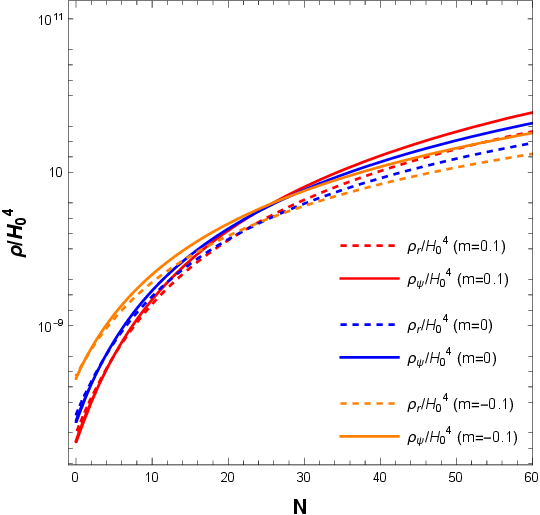}
	\caption{Evolution of the dimensionless inflaton energy densities (solid line), and of the dimensionless radiation energy densities (dashed line) with e-folds number $N$ for different values of $m$: $m=0.1$ (red), $m=0$ (blue), $m=-0.1$ (orange), in the model $\mathcal{L}_\omega=-\eta H_0^{4-2n-m}\psi^m\left(-\omega_\mu\omega^\mu\right)^n$. $N$ evolves from a large value up to $N=0$ where inflation ends. Other parameters are: $\eta=0.5, \gamma=72, n=1.5$.}
	\label{2energy}
\end{figure}

Fig.~\ref{2energy} shows the time evolution of the dimensionless inflaton energy density $\rho_\psi/H_0^4$ and the dimensionless radiation energy density $\rho_r/H_0^4$ for different functional forms of $\psi$ in the $\mathcal{L}_\omega$ term. The value of $m$ significantly influences both the inflaton energy density and the radiation density during inflation. For different values of the parameter $m$, the system consistently exhibits a process where radiation overtakes the inflaton, and the crossing occurs around $N=4$. At the end of inflation, compared with the minimal coupling model, a positive $m$ yields a lower radiation density, whereas a negative $m$ yields a higher one.

\begin{figure}[t]
	\includegraphics[width=8.5cm,height=8.5cm]{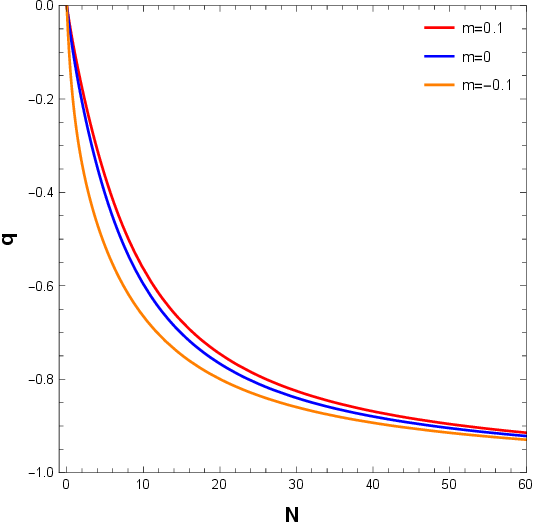}
	\caption{Variations as functions of the e-folds number $N$ of the deceleration parameter $q$ in the model $\mathcal{L}_\omega=-\eta H_0^{4-2n-m}\psi^m\left(-\omega_\mu\omega^\mu\right)^n$ for different $m$. The values of $m$ are: $m=0.1$ (red), $m=0$ (blue), $m=-0.1$ (orange). $N$ evolves from a large value up to $N=0$ where inflation ends. Other parameters are: $\gamma=0.72, \eta=0.5, n=1.5$.}
	\label{2q}
\end{figure}

The variations of deceleration parameter $q$ with respect to $N$ are illustrated in Fig. \ref{2q}. Similar to the minimal coupling scenario, the deceleration parameter $q$ monotonically rises from negative to positive, marking a natural transition from an inflationary de Sitter universe to a decelerating, radiation-dominated universe, across different forms of $\mathcal{L}_\omega$. The effect of $m$ on the evolution of $q$ is much more significant than that of the parameter $\gamma$.

The evolution of $Q$ and $T_r/H$ as functions of $N$ for various $m$ values is displayed in Fig. \ref{2QTH}. Both $Q$ and $T_r/H$ increase as $N$ decreases. {Compared with the minimal coupling model, when $m > 0$, the evolution of $Q$ and $T_r/H$ is almost the same as that in the minimal coupling model, while when $m < 0$, $Q$ and $T_r/H$ are significantly reduced. Similar to the minimal coupling scenario, we still have $T_r/H<1$ for $N>40$, indicating that radiation fluctuations do not significantly dominate over quantum fluctuations.

\begin{figure*}[t]
	\begin{center}
		\begin{subfigure}{0.495\textwidth}
			\includegraphics[width=\textwidth, height=0.7\textwidth]{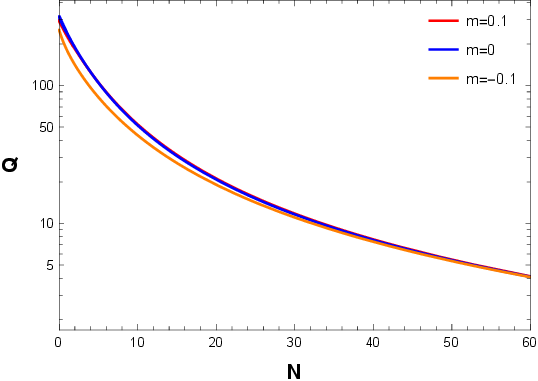}
		\end{subfigure}
		\hfill
		\begin{subfigure}{0.495\textwidth}
			\includegraphics[width=\textwidth, height=0.7\textwidth]{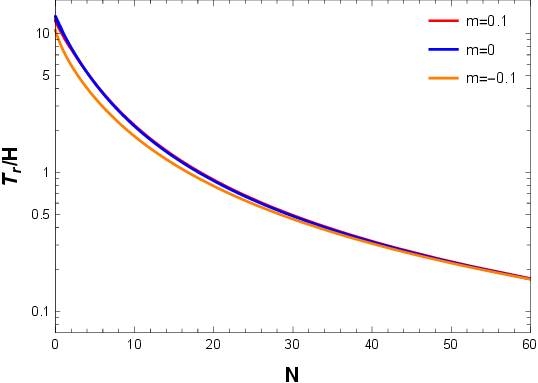}
		\end{subfigure}
	\end{center}
	\caption{Variations as functions of e-folds number $N$ of the dissipation ratio $Q$ (left panel) and of the thermal ratio $T_r/H$ (right panel) in the model $\mathcal{L}_\omega=-\eta H_0^{4-2n-m}\psi^m\left(-\omega_\mu\omega^\mu\right)^n$. The values of $m$ are: $m=0.1$ (red), $m=0$ (blue), $m=-0.1$ (orange). $N$ evolves from a large value up to $N=0$ where inflation ends. Other parameters are: $\gamma=72, \eta=0.5, n=1.5$.}
	\label{2QTH}
\end{figure*}

The evolution of the radiation energy density and temperature under different $\mathcal{L}_\omega$ forms is presented in Fig. \ref{2rT}. The exponent $m$ significantly influences their evolution. For larger values of $N$ ($N>40$), $m>0$ corresponds to a larger radiation density, while $m<0$ corresponds to a smaller one. This trend reverses at a certain $N$, for $N<20$, $m>0$ gives a smaller radiation density and $m<0$ a larger one. Temperature follows a similar trend: $m<0$ gives larger $\theta_\text{end}$, $m>0$ smaller. For different $m$, the temperatures at the end of inflation $\theta_\text{end}$ are given in Table \ref{tab3}.

\begin{figure*}[t]
	\begin{center}
	\begin{subfigure}{0.495\textwidth}
		\includegraphics[width=\textwidth, height=0.7\textwidth]{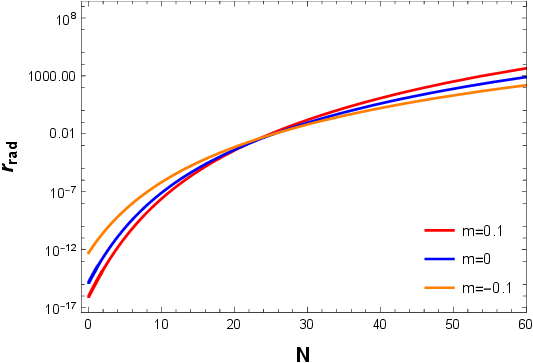}
	\end{subfigure}
	\hfill
	\begin{subfigure}{0.495\textwidth}
		\includegraphics[width=\textwidth, height=0.7\textwidth]{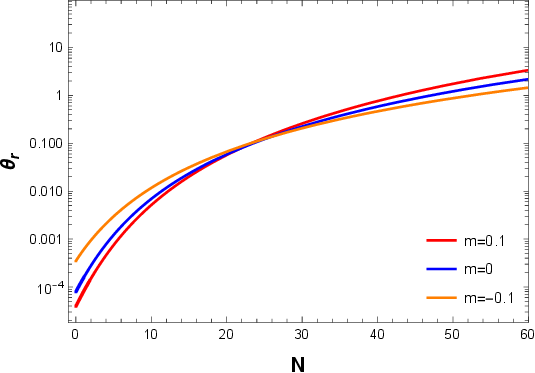}
	\end{subfigure}
	\end{center}
	\caption{Variations as functions of e-folds number $N$ of the dimensionless radiation density $r_\text{rad}$ (left panel) and of the dimensionless temperature $\theta_r$ (right panel) in the model $\mathcal{L}_\omega=-\eta H_0^{4-2n-m}\psi^m\left(-\omega_\mu\omega^\mu\right)^n$. The values of $m$ are: $m=0.1$ (red), $m=0$ (blue), $m=-0.1$ (orange). $N$ evolves from a large value up to $N=0$ where inflation ends. Other parameters are: $\gamma=72, \eta=0.5, n=1.5$.}
	\label{2rT}
\end{figure*}

\begin{table}[htbp]
	\begin{center}
		\begin{tabular*}{0.5\columnwidth}{@{\extracolsep{\fill}}c c@{\extracolsep{\fill}}}
			\hline
			\textbf{$m$} & \textbf{$\theta_\text{end} (10^{-5})$}  \\
			\hline
			$0.1$ & $3.94$  \\
			$0$ & $7.91$  \\
			$-0.1$ & $34.86$  \\
			\hline
		\end{tabular*}
	\end{center}
	\caption{The temperatures at the end of inflation $\theta_\text{end}$ for different values of $m$, in the model $\mathcal{L}_\omega=-\eta H_0^{4-2n-m}\psi^m\left(-\omega_\mu\omega^\mu\right)^n$. Other parameters are: $\eta=0.5, \gamma=72, n=1.5$.}
	\label{tab3}
\end{table}

Fig. \ref{2nsrN} shows the values of scalar spectral index $n_s$ (left panel) and tensor-to-scalar ratio $r$ (right panel) with respect to number of e-folds $N$ for different $m$. $n_s$ increases as $N$ decreases, while $r$ decreases with decreasing $N$. Unlike the minimal coupling model, $\mathcal{L}_\omega$ directly enters the evolution equation of the inflaton Eq. \eqref{inflatonequation}. Thus, the functional form of $\psi$ in $\mathcal{L}_\omega$ can be regarded as a modification to the inflaton potential $V(\psi)$, thereby significantly influencing the observational predictions. As can be seen from Fig. \eqref{2nsrN}, compared with the minimal coupling model, the model with positive $m$ yields a higher $n_s$ and a lower $r$. Conversely, the model with negative $m$ produces a lower $n_s$ and a higher $r$.

\begin{figure*}[t]
	\begin{center}
		\begin{subfigure}{0.495\textwidth}
			\includegraphics[width=\textwidth, height=0.7\textwidth]{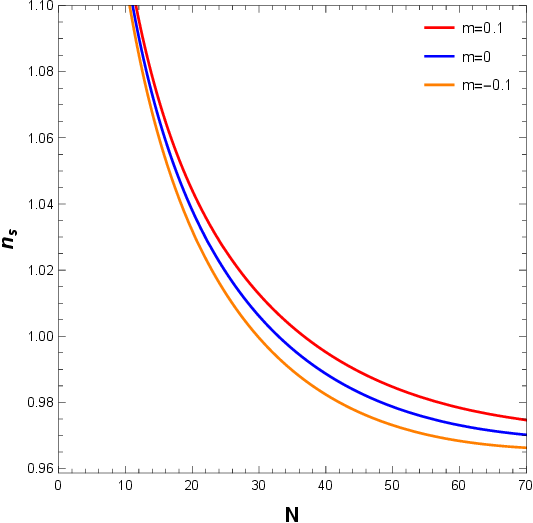}
		\end{subfigure}
		\hfill
		\begin{subfigure}{0.495\textwidth}
			\includegraphics[width=\textwidth, height=0.7\textwidth]{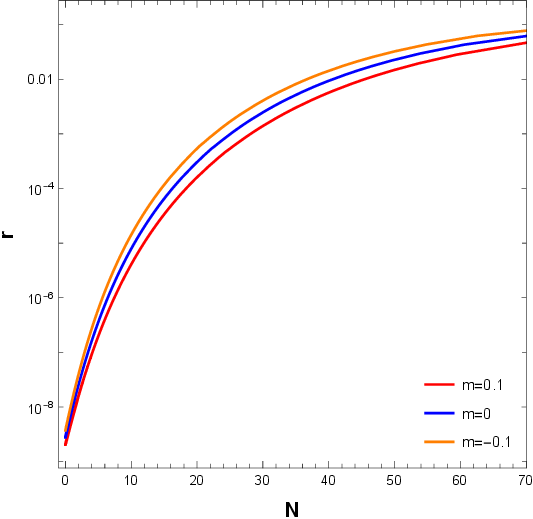}
		\end{subfigure}
	\end{center}
	\caption{The values of scalar spectral index $n_s$ (left panel) and tensor-to-scalar ratio $r$ (right panel) with respect to number of e-folds $N$ for different parameters: $m=0.1$ (red), $m=0$ (blue), $m=-0.1$ (orange), in the model $\mathcal{L}_\omega=-\eta H_0^{4-2n-m}\psi^m\left(-\omega_\mu\omega^\mu\right)^n$. $N$ evolves from a large value up to $N=0$ where inflation ends. Other parameters are: $\gamma=72, \eta=0.5, n=1.5$.}
	\label{2nsrN}
\end{figure*}

Fig. \ref{2rations} shows the tensor-to-scalar ratio $r$ against the scalar spectral index $n_s$ for different values of $m, N_*$. For different values of $m$, the curves show similar behavior. $r$ decreases as $n_s$ increases. Increasing the value of $m$ shifts the curve toward the upper‑right region, until a certain maximum $m$ is reached, beyond which the curve no longer passes through the allowed region. On each curve, $\gamma$ increases from left to right. Each value of $m$ and $N_*$ corresponds to a specific range of $\gamma$ that satisfies the observational constraints. The correspondence between $m, N_*$ and the viable $\gamma$ ranges is provided in Table \ref{tab2N}.

\begin{figure}[t!]
	\includegraphics[width=8.5cm, height=8.5cm]{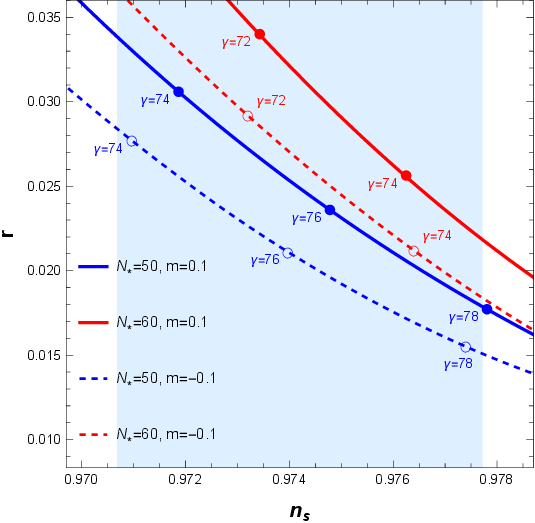}
	\caption{The tensor-to-scalar ratio $r$ against the scalar spectral index $n_s$ for different values of $m, N$: $m=0.1, N_*=50$ (blue solid), $m=0.1, N_*=60$ (red solid), $m=-0.1, N_*=50$ (blue dashed), $m=-0.1, N_*=60$ (red dashed), in the model $\mathcal{L}_\omega=-\eta H_0^{4-2n-m}\psi^m\left(-\omega_\mu\omega^\mu\right)^n$. On each curve, $\gamma$ increases from left to right. The blue region denotes the parameter space consistent with the constraints from Eqs. \eqref{nscons} and \eqref{ratiocons}. $N$ correspond to CMB pivot scale and the values of $\kappa$ are chosen to enforce $\mathcal{P}_R\simeq2.1\times10^{-9}$. Other parameters are: $\eta=0.5, n=1.5$.}
	\label{2rations}
\end{figure}

\begin{table}[htbp]
	\begin{center}
		\begin{tabular*}{0.5\columnwidth}{@{\extracolsep{\fill}}c c c@{\extracolsep{\fill}}}
			\hline
			\textbf{$N_*$} & \textbf{$m$} & \textbf{$\gamma$}  \\
			\hline
			$50$ & $0.1$ &  $73.23 - 77.93$ \\
			$60$ & $0.1$ &  $71.50 - 75.76$ \\
			$50$ & $-0.1$ & $73.83 - 78.07$ \\
			$60$ & $-0.1$ & $70.12 - 74.69$ \\
			\hline
		\end{tabular*}
	\end{center}
	\caption{The suitable ranges of $\gamma$ corresponding to different values of $m, N_*$, in the model $\mathcal{L}_\omega=\eta H_0^2\omega_\mu\omega^\mu$. Other parameters are: $\eta=0.5, \gamma=72, n=1.5$.}
	\label{tab2N}
\end{table}

\section{Discussions and final remarks}\label{5}

In the present paper, we have investigated the dynamics of warm inflation within the framework of a specific physical model constructed in the framework of Weyl geometry. Our starting point is  the Weyl $\tilde{R}^2$ theory, which we have extended by  incorporating matter fields in a conformally invariant way. Hence, the conformal invariance of the theory is fully preserved.  The theory is equivalent to a linear gravitational theory in the Ricci scalar, with an additional scalar degree of freedom introduced, which originates from the non-linear curvature term through its linearization. By taking into account the relation between $R$ and $\tilde{R}$, the action of the theory can be formulated in a Riemannian space. The gravitational field equation, the vector field  equation and the scalar equation, all formulated in the Riemannian space, have also been derived. The gravitational field equations have been derived by using the metric variation of the action. Furthermore, we have shown that in this model, in order to maintain the full conformal invariance of the action, the effective matter Lagrangian must include the Weyl vector, thereby introducing a novel form of coupling between the Weyl vector and the auxiliary scalar field.

We have investigated the cosmological implications and tests of the Weyl geometric gravity theory by considering the warm inflationary scenario of the early evolution of the Universe. We have derived the cosmological evolution equations (the generalized Friedmann equations) for the theory in a flat Friedmann-Lemaitre-Robertson-Walker (FLRW) Universe. The four physical components present in the theory: the scalar field $\phi$, the inflaton field $\psi$, the Weyl vector $\omega_\mu$, and the radiation field $\rho_r$, coexist in the Universe, and collectively govern its dynamical evolution. We have obtained the energy balance equation for the theory, which includes an additional term contributed by $\mathcal{L}_\omega$. The starting point in the investigation  of the radiation creation is the generalized energy balance equation (\ref{energy}), formulated in the FLRW geometry, which due to the presence of an extra (source) term can be naturally interpreted as describing the energy transfer from the Weyl and scalar fields to some forms of matter. Similarly to the standard warm inflationary scenario the energy balance equation can be split to describe radiation production during the very early stages of the cosmological evolution.

Based on the dependence of $\mathcal{L}_\omega$ on the function $\psi$, three types of models can be formulated: the minimal coupling model ($m=0$), the sourcing model ($m>0$), and the dissipative model ($m<0$). We performed numerical simulations for each model, with the widely studied dissipation coefficient $\Gamma=\gamma T_r$, and the inflaton potential $V(\psi)=\kappa\psi^4$.

For the minimal coupling model, we have studied the evolutionary curves of the radiation density and the inflaton energy density, and we have obtained the characteristics of the warm inflation process in which the Universe transitions from inflaton-dominated to radiation-dominated expansion. In addition, we present the temporal evolution of the deceleration parameter, radiation energy density, and temperature, demonstrating their explicit dependence on the parameter $\eta$. 

The model's initial Hubble parameter $H_0$ is related to the reheating temperature through the dimensionless temperature at the end of inflation $\theta_\text{end}$. In order to compare the model with the observations we have also investigated the variations  of the scalar spectral index $n_s$ and of the tensor-to-scalar ratio $r$ against the e-folding number $N$, by obtaining the dependence of $n_s$ and $r$ on the model parameters $m$ and $\gamma$. 

In the minimal coupling model, the observables are significantly influenced by $\gamma$, while remaining largely insensitive to $\eta$. The scalar spectral index $n_s$ increases with $\gamma$, while the tensor-to-scalar ratio $r$ decreases with $\gamma$. For $N=50 - 60$, the allowed values of $\gamma$ lies within $\gamma=70.67 - 78.17$. Typically, the allowed range shifts toward lower $\gamma$ values as $N$ increases.

The non-minimal coupling models also successfully describe warm inflation scenarios. We have obtained the evolution of the deceleration parameter $q$, of the radiation energy density $\rho_r$, and of the  temperature $\theta_r$ for different values of the exponent $m$. While they follow patterns similar to those in the minimal coupling case, key features such as the temperature at the end of inflation are significantly influenced by the specific value of $m$. The observable quantities are also influenced by the value of $m$. The scalar spectral index $n_s$ increases with $m$, while the tensor-to-scalar ratio $r$ decreases with $m$. The combined constraints from $r, n_s$, and $N$ jointly restrict the allowed range of the parameters $\gamma$ and $m$.

It is interesting to note that in the present model both the scalar field and the Weyl vector do increase during the early Universe evolution, indicating the presence of a complex dynamical behavior. The presence of these quantities in the late Universe may play an important role in the recent cosmological dynamics, mimicking, for example, an effective cosmological constant, and thus triggering the accelerated expansion of the Universe.

The creation of a radiation fluid, as well as an early accelerated expansion have very important implications not only for the dynamics and evolution of the very early Universe, but also for the present day composition and structure of the cosmological environment. In the approach considered in the present work these processes are determined by the scalar field and by the Weyl vector, interacting with the radiation fluid. The details of the interaction also essentially depend on the physical parameters describing this interaction, as well as and on the parameters of the quantum particle physics models necessary to describe the physical processes related to the creation of the new particles. However, presently there is no physical theory providing, for example, the numerical values of the physical parameters describing, the interaction between the Weyl vector or a scalar field and a radiation fluid. In the present work we have investigated a geometric framework for the description of the radiation fluid production processes from two scalar fields and from the Weyl vector, having a geometric origin, and intimately related to the structure of the space-time manifold. The obtained results represent some preliminary theoretical investigations that could lead to the deeper understanding of the complex physical and geometrical processes that took place and shaped the early Universe, with effects extending to the present time.

\section*{Acknowledgments}

We would like to thank the anonymous reviewer for comments and suggestions that helped us to significantly improve our manuscript. This work was supported in part by the National Natural Science Foundation of China (NSFC) under Grant No. 12275367, the Fundamental Research Funds for the Central Universities, and the Sun Yat-Sen University Science Foundation.

\appendix

\section{Derivation of energy balance equation}\label{A}

From Eq. \eqref{vec}, we have
\begin{equation}\label{A1}
	-2\omega^2\frac{\partial\mathcal{L}_\omega}{\partial\omega^2}=-\frac{\omega\phi^2}{4\xi^2}\left(\omega-2\frac{\dot{\phi}}{\phi}\right).
\end{equation}
Insert Eqs. \eqref{rw} and \eqref{A1} into Eq. \eqref{fir}, we have
\begin{equation}\label{A2}
	3H^2=-6H\frac{\dot{\phi}}{\phi}+\frac{1}{4}\phi^2-\frac{3}{4}\omega^2+\frac{6\xi^2}{\phi^2}\left(\rho_m-\mathcal{L}_\omega\right).
\end{equation}
Multiply Eq. \eqref{A2} by $\phi^2$
\begin{equation}\label{A3}
	3H^2\phi^2=-6H\dot{\phi}\phi+\frac{1}{4}\phi^4-\frac{3}{4}\omega^2\phi^2+6\xi^2\left(\rho_m-\mathcal{L}_\omega\right).
\end{equation}
Take the time derivative of both sides of Eq. \eqref{A3}
\begin{align}\label{A4}
	6\dot{H}H\phi^2+6H^2\dot{\phi}\phi=
	&-6\dot{H}\dot{\phi}\phi-6H\ddot{\phi}\phi-6H\dot{\phi}^2+\phi^3\dot{\phi}\nonumber\\
	&-\frac{3}{2}\dot{\omega}\omega\phi^2-\frac{3}{2}\omega^2\dot{\phi}\phi\nonumber\\
	&+6\xi^2\left(\dot{\rho}_m-\dot{\mathcal{L}}_\omega\right).
\end{align}
Insert Eq. \eqref{A2} into Eq. \eqref{sec}
\begin{align}\label{A5'}
	2\dot{H}=-2\frac{\ddot{\phi}}{\phi}-2\frac{\dot{\phi}^2}{\phi^2}+2H\frac{\dot{\phi}}{\phi}+3\omega\frac{\dot{\phi}}{\phi}-\frac{6\xi^2}{\phi^2}\left(\rho_m+p_m\right).
\end{align}
Multiply Eq. \eqref{A5'} by $3H\phi^2$
\begin{align}\label{A5}
	6H\dot{H}\phi^2
	=&-6H\ddot{\phi}\phi-6H\dot{\phi}^2+6H^2\dot{\phi}\phi+9H\omega\dot{\phi}\phi \nonumber\\
	&-18\xi^2H\left(\rho_m+p_m\right).
\end{align}
Insert Eq. \eqref{A5} into \eqref{A4}, we obtain
\begin{align}\label{A6}
	6\xi^2\left[\dot{\rho}_m+3H\left(\rho_m+p_m\right)\right]=
	&12H^2\dot{\phi}\phi-\phi^3\dot{\phi}+6\dot{H}\dot{\phi}\phi \nonumber\\
	&+9H\omega\dot{\phi}\phi+\frac{3}{2}\omega^2\dot{\phi}\phi+3\dot{\omega}\dot{\phi}\phi \nonumber\\
	&-3\dot{\omega}\dot{\phi}\phi+\frac{3}{2}\dot{\omega}\omega\phi^2 \nonumber\\
	&+6\xi^2\dot{\mathcal{L}}_\omega.
\end{align}
According to Eqs. \eqref{sca} and \eqref{vec}, the second and the third lines of Eq. \eqref{A6} are
\begin{align}\label{A7}
	\text{The second line}
	&=9H\omega\dot{\phi}\phi+\frac{3}{2}\omega^2\dot{\phi}\phi+3\dot{\omega}\dot{\phi}\phi\\\nonumber
	&=-12H^2\dot{\phi}\phi+\phi^3\dot{\phi}-6\dot{H}\dot{\phi}\phi,
\end{align}
\begin{align}\label{A8}
	\text{The third line}
	&=-3\dot{\omega}\dot{\phi}\phi+\frac{3}{2}\dot{\omega}\omega\phi^2\\\nonumber
	&=\frac{3}{2}\dot{\omega}\phi^2\left(\omega-2\frac{\dot{\phi}}{\phi}\right)\\\nonumber
	&=12\xi^2\omega\dot{\omega}\frac{\partial\mathcal{L}_\omega}{\partial\omega^2}.
\end{align}
$\dot{\mathcal{L}}_\omega$ can be written as
\begin{equation}\label{A9}
	\dot{\mathcal{L}}_\omega=-2\omega\dot{\omega}\frac{\partial\mathcal{L}_\omega}{\partial\omega^2}+\dot{\psi}\frac{\partial\mathcal{L}_\omega}{\partial\psi}.
\end{equation}
Insert Eqs. \eqref{A7}, \eqref{A8} and \eqref{A9} into Eq. \eqref{A6}, we obtain
\begin{equation}
	\dot{\rho}_m+3H\left(\rho_m+p_m\right)=\dot{\psi}\frac{\partial\mathcal{L}_\omega}{\partial\psi}.
\end{equation}

\bibliography{refs}

\end{document}